\documentstyle[12pt]{article}

\newcommand \beq{\begin{eqnarray}}
\newcommand \eeq{\end{eqnarray}}

\def\frac#1#2{{#1 \over #2}}
\def\smallfrac#1#2{{\textstyle {#1 \over #2}}}

\def\simge{\mathrel{%
   \rlap{\raise 0.511ex \hbox{$>$}}{\lower 0.511ex \hbox{$\sim$}}}}
\def\simle{\mathrel{
   \rlap{\raise 0.511ex \hbox{$<$}}{\lower 0.511ex \hbox{$\sim$}}}}

\def\slashchar#1{\setbox0=\hbox{$#1$}           
   \dimen0=\wd0                                 
   \setbox1=\hbox{/} \dimen1=\wd1               
   \ifdim\dimen0>\dimen1                        
      \rlap{\hbox to \dimen0{\hfil/\hfil}}      
      #1                                        
   \else                                        
      \rlap{\hbox to \dimen1{\hfil$#1$\hfil}}   
      /                                         
   \fi}                                         %

\def\subrightarrow#1{
  \setbox0=\hbox{
    $\displaystyle\mathop{}
    \limits_{#1}$}
  \dimen0=\wd0
  \advance \dimen0 by .5em
  \mathrel{
    \mathop{\hbox to \dimen0{\rightarrowfill}}
       \limits_{#1}}}                           

\def\subleftarrow#1{
  \setbox0=\hbox{
    $\displaystyle\mathop{}
    \limits_{#1}$}
  \dimen0=\wd0
  \advance \dimen0 by .5em
  \mathrel{
    \mathop{\hbox to \dimen0{\leftarrowfill}}
       \limits_{#1}}}                           

\def\journal#1#2#3#4{\ {#1}{\bf #2} ({#3})\  {#4}}

\def\NPB{\journal{Nucl.\ Phys.\ {\bf B}}}

\def\PRD{\journal{Phys.\ Rev.\ {\bf D}}}

\begin{document}

\def\funcderv#1#2{\frac{\delta #1}{\delta #2}}

\newcommand\YDW{\mbox{$\dot{y}^{\alpha}\nabla_{\alpha}W_{\mu}  +
 W_{\alpha}\nabla_{\mu}\dot{y}^{\alpha} -
 \dot{y}^{\alpha}W_{\alpha}\Lambda,_{\mu}$}}

\newcommand\YDMM{\mbox{$(\dot{y}^{\alpha}\nabla_{\alpha}m^{\mu}  -
m^{\lambda}\nabla_{\lambda}\dot{y}^{\mu} +
 \dot{y}^{\mu} m^{\lambda}\partial_{\lambda} \Lambda ) $}}

\newcommand\YDMN{\mbox{$(\dot{y}^{\alpha}\nabla_{\alpha}m^{\nu}  -
m^{\lambda}\nabla_{\lambda}\dot{y}^{\nu} +
\dot{y}^{\nu} m^{\lambda}\partial_{\lambda} \Lambda) $}}

\newcommand\YFETAM{\mbox{$\eta_{\rho}^{\nu}m^{\rho}\dot{y}^{\lambda}
F_{\nu\lambda}$}}

\newcommand\AMBIGUITYM{\mbox{$ ( m_{\rho} \nabla ^{\nu} \dot{y}
^{\rho} - m \cdot \dot {y} \Lambda,^{\nu} + m^{\nu} \dot{y} \cdot
\nabla \Lambda + m \cdot \nabla \dot{y}^{\nu} - \dot{y}^{\nu} m \cdot
\nabla \Lambda )$}}

\newcommand\measM{\mbox{$ d^4x (\sqrt{-g} e^{\Lambda}) $}}
\newcommand\WW{\mbox{$\frac{3}{4}m_{g}^{2}W_{\mu }W_{\nu }$}}
\newcommand\WWW{\mbox{$\frac{3}{4}m_{g}^{2}W_{0\mu }W_{0\nu }$}}
\newcommand\WCW{\mbox{$\frac{3}{4}m_{g}^{2}W_{0}^{\mu}W_{0 \mu}$}}
\newcommand\WAWB{\mbox{$\frac{3}{4}m_{g}^{2}W_{0\alpha}W_{0\beta}$}}
\newcommand\YVF{\mbox{$V^{\nu}\dot{y}^{\lambda}F_{\nu\lambda}$}}
\newcommand\amb{\mbox{$V^{\nu}\dot{y}^{\lambda}F_{\nu\lambda}$}}
\newcommand\meas{\mbox{$ d^4x (\sqrt{-g} g^{\mu\nu} e^{\Lambda}) $}}
\newcommand\ambig{\mbox{$W_{\lambda}\nabla_{\mu}\dot{y}^{\lambda}-\dot{y}
\cdot W \partial \Lambda_{\mu} +g_{\mu\lambda} W \cdot \nabla
\dot{y}^{\lambda} + W_{\mu} \dot{y} \cdot \partial \Lambda
 -\dot{y}_{\mu}W_{\alpha} \Lambda,^{\alpha}$}}

\newcommand\KETA{\mbox{$[Q_{\alpha} Q_{\beta} K^{2}
+\eta_{\alpha\beta} - Q_{\{ \alpha} K_{\beta \}}] $}}

\newcommand\KYDOT{\mbox{$[Q_{ \{ \alpha} \eta_{\beta \} }^{\lambda}
-Q_{\alpha} Q_{\beta} K^{\lambda}] $}}

\newcommand\AMBIGUITY{\mbox{$[(Q^{\nu}P_{\alpha}-\eta^{\nu}_{\alpha})
(Q^{\mu}R_{\beta}
- \eta^{\mu}_{\beta}) + (Q^{\mu}P_{\beta} -\eta^{\nu}_{\beta})
(Q^{\mu}R_{\alpha} - \eta^{\mu}_{\alpha}) ]$}}

\newcommand\AMBIGUITYY{\mbox{$- \KYDOT K^{\rho}[(Q^{\nu}P_{\rho} -
\eta^{\nu}_{\rho})(Q^{\mu}R_{\lambda} - \eta^{\mu}_{\lambda}) +
(P,\nu) \rightleftharpoons (R,\mu)]$}}
\newcommand\AMBIGUITYYY{\mbox{$- \KETA [(Q^{\nu}P_{\rho} -
\eta^{\nu}_{\rho})(Q^{\mu}R^{\rho} - \eta^{\mu \rho})]$}}

\newcommand\ghtl{\mbox{$\Gamma^{HTL}$}}
\newcommand\EMLAM{\mbox{$[Q_{\alpha} Q_{\beta} \Box (Q\cdot \partial )^{-2}-
 Q_{\{ \alpha}\partial_{\beta \}}(Q\cdot \partial )^{-1}] $}}
\newcommand\EMYDOT{\mbox{$[Q_{ \{ \alpha} \eta_{\beta \} }^{\lambda} -
Q_{\alpha} Q_{\beta}(Q\cdot \partial )^{-1}\partial^{\lambda}] $}}

\newcommand\EMETA{\mbox{$[Q_{\alpha} Q_{\beta} \Box (Q\cdot \partial )^{-2}
+\eta_{\alpha\beta} - Q_{\{ \alpha}\partial_{\beta \}}
(Q\cdot \partial )^{-1}] $}}

\newcommand\yslash{\mbox{$e^{-\frac{\Lambda}{2}} \dot{\slashchar{y}  }$ }  }
\newcommand\yd{\mbox{$e^{\Lambda} \dot{y} \cdot \tilde{\nabla}$ }}

\newcommand\pbzero{\mbox{$\bar{\Psi_{0}}$}}
\newcommand\pzero{\mbox{$\Psi_{0}$}}
\newcommand\spp{\hspace{0.3cm}}
\newcommand\sppp{\hspace{0.1cm}}

\begin{titlepage}
\begin{flushright}
{DAMTP 94--74}
\end{flushright}
\vspace*{2cm}
\begin{center}
\baselineskip=13pt
{\Large  Hard Thermal Loops, Weak Gravitational Fields and
 The Quark Gluon Plasma Energy Momentum Tensor. \\}
\vskip0.5cm
Eamonn Gaffney \\ {\em DAMTP, University of Cambridge, UK}
\vskip0.5cm
September 1994
\end{center}

\vskip 2cm
\begin{abstract}
We use an auxiliary field construction to discuss the hard thermal
 loop effective action associated with
  massless thermal SU(N) QCD  interacting with a weak gravitational
field. It is  demonstrated that
the previous attempt to derive this effective
action has only been partially successful and that it is presently
only known to  first order in the graviton coupling
constant. This is still sufficient to enable a calculation of a
symmetric traceless quark
gluon plasma energy momentum tensor.
Finally, we comment on the conserved currents and charges of
the derived energy momentum tensor.
 \end{abstract}

\vskip 3cm


\end{titlepage}

\section{Introduction}

There has been much discussion on various ways of deriving the energy
momentum tensor of the high temperature
deconfined massless quark gluon plasma (QGP), which is  non-standard due to
the non-local nature of the perturbative resummation required to
describe the plasma even at the lowest orders of $g$, the QCD coupling
constant  \cite{brateen}. Weldon \cite{Weldon} describes the non-local
 dynamics of the QGP using a lagrangian with a
local dependence on auxiliary fields (rather than a non local
dependence on physical fields). He then proceeds to derive the QGP energy
momentum tensor by calculating
the canonical energy momentum tensor of this
lagrangian (which however is neither symmetric nor traceless despite
the fact that the (classical) field theory of the massless QGP is scale
invariant).
Blaizot and Iancu \cite{bi} deduce three forms of the QGP
energy momentum tensor for a pure gluon plasma by integrating the
divergence condition $\partial_{\nu}T^{\mu\nu} = j^{\mu}$ (where
$j^{\mu}$ is a current due to an external source).
 This approach however does not
display the link between symmetries of the theory and the
conserved quantities constructed from the energy momentum tensor.

Brandt, Frenkel and Taylor \cite{taylor} discuss the interaction of the
QGP with a weak gravitational field and introduce  the two
effective actions $\Gamma^{HTL}[A, g]$ and $\Gamma^{HTL(quark)}[A,
e]$, the former for gluon graviton interactions
 and the latter for the interaction of quarks with gluons and
 vierbien fields  (where $A$, $g$, $e$ are (classical) gluon, metric, and
vierbien fields).
They  define the QGP energy momentum tensor to be the sum of the two
 expressions
\begin{eqnarray*}
     T^{HTL}_{\alpha \beta} \equiv ~2lim_{g \rightarrow \eta} \Big(
               \funcderv{\Gamma^{HTL} [A, g]}{g^{\alpha
\beta}} \Big) && T^{HTL(quark)}_{\alpha \beta}
\equiv ~lim_{e \rightarrow \eta} \Big(
     ~~e^{a}_{~\alpha} \funcderv{\Gamma^{HTL(quark)}[A, e]}{e^{a
\beta}} \Big)
\end{eqnarray*}
which are  the definitions of $T^{HTL}_{\alpha \beta}$ and
$T^{HTL(quark)}_{\alpha \beta}$ adopted in this
paper. Note that the above definition of $T^{HTL}_{\alpha \beta}$
 is equivalent to  twice the leading temperature
contribution of the sum over $n$ of thermal Feynman graphs consisting of
 a graviton
current insertion $ T_{\alpha \beta} ^{ \{ T=0 \} }$ and
  $n$ gluon current insertions. ($T_{\alpha \beta}^{ \{T=0 \}}$ is the
 usual gluon contribution to the zero temperature QCD energy momentum
tensor. It  is also the zero temperature coupling to the graviton field,
$\varphi^{\alpha \beta}$, which is related to the metric by
 $g^{\alpha \beta} = \eta^{\alpha \beta} + \kappa
\varphi^{\alpha \beta}$ where $\kappa$ is the graviton coupling
constant). A similar
interpretation can be given for the quark tensor.

 The properties of the energy momentum tensor such as
divergencelessness (unlike Blaizot and Iancu  we do not have external
sources), tracelessness and symmetry (on use of the equations of motion
if need be) are directly linked to the symmetries of the hard thermal
loop theory. These properties  can be deduced either from the
invariances of  the above effective
actions  or, equivalently, from the properties of hard
thermal loop diagrams with graviton insertions.

In this paper the two effective actions above are written, to first order
 in the graviton coupling constant
 $\kappa \ll 1$, in
terms of  local actions with auxiliary fields. This greatly simplifies
the calculation of $T^{HTL}_{\alpha \beta}$ and
$T^{HTL(quark)}_{\alpha \beta}$. We note that a traceless and
symmetric energy momentum tensor including a contribution due to  the
presence of quarks has not previously been written to all orders in
$g$ in the literature. Furthermore, the auxiliary field
construction explicity shows that the gluon hard thermal loop effective
action (with no gravitons), together with the symmetry properties of the
gluon graviton
 effective actions given in \cite{taylor} (and by conditions
(A) $\rightarrow$ (D) of section (2.2) below), are {\em not} sufficient to
determine the gluon graviton effective action $\Gamma^{HTL}[A,
g]$, even at $O(\kappa)$. This  is contrary to
assertion made in  \cite{taylor}. We give supplementary conditions
that enable determination of $\Gamma^{HTL}[A, g]$ to
$O(\kappa)$. Further, it is
 shown that both  $\Gamma^{HTL}[A, g]$ and
$\Gamma^{HTL(quark)}[A, g]$ are not known at present to $O(\kappa^2)$.

We show that the derived gluon contribution to the energy momentum
tensor, $T^{HTL}_{\alpha \beta}$, gives an
integrated energy density (i.e. $P_{0} = \int d^3x T^{HTL}_{00}$) which
is a positive
definite functional of the gauge fields, although this property is
not manifest. This is a straightforward corollary of the results of Blaizot
and Iancu \cite{bi}. Finally, it is  shown that the gluon contribution to
the integrated energy density
is also given by the Hamiltonian associated with the flatspace local
auxiliary field action on  elimination of  the auxiliary fields.

\section{Gluon fields}
\subsection{Flat space Gluon auxiliary fields}

The notation used in this paper is as follows: we consider the
interaction of a weak external graviton field with massless thermal
SU(N) QCD in the deconfined phase at temperature T and  zero chemical
potential with   $N_f$
quark flavours. The generators of the fundamental and adjoint
representations of the colour group are denoted respectively by $t^A$
and $T^A$. These are taken to be hermitian and are normalised by
\beq
tr(t^At^B)= \frac{1}{2} \delta^{AB}  \qquad tr(T^AT^B)= N \delta^{AB}
\eeq
The structure constants $f^{ABC}$ satisfy $[t^A, t^B] = + i f^{ABC} t^C$
and $(T^A)_{BC} = - i f^{ABC} $.
We have quark fields $\psi$, gauge fields in the two above
representations $A_{\mu}=A_{\mu}^{A}t^{A}$
${\cal A}_{\mu}=A_{\mu}^{A}T^{A}$, vierbien fields $e_{a \mu}$ where
$a$ denotes a  local Lorentz index and graviton fields
$\phi^{\alpha \beta}$ given by
 \beq
g^{\alpha\beta}=\eta^{\alpha\beta}+\kappa\varphi^{\alpha\beta}
\eeq
where $\kappa\ll1$.

Initially we deal with the gluon sector of the theory only, although the
inclusion of quarks will be considered later.  We first discuss
section (3) of Weldon's paper \cite{Weldon}. Consider
introducing an action with
auxiliary fields $V_{0}^{\mu}(Q, x) = V_{0}^{\mu A}(Q, x) t^A$ and
$W_{0}^{\mu}(Q, x) = W_{0}^{\mu A}(Q, x)t^A $ which transform
like the field strength under gauge transformations.
Here $Q$ is a light-like vector of the form $(1, \bf{\hat{q}})$ and the
measure $d\Omega (4 \pi)^{-1}$  used below averages over all possible
$\bf{\hat{q}}$ and the label
$_{0}$ indicates flat space. Weldon \cite{Weldon} defines the action
$\Gamma_0$ by
\beq  \nonumber
g^2T^2 X_{0}(Q;V_0, W_0, A, \eta] &=&  \int
\frac{d\Omega}{4 \pi} d^{4}x \/ 2 \/ \mbox{tr } [ -\WWW +
V_{0}^{\mu}Q^{\nu}F_{\mu\nu} - V_{0}^{\mu}Q \cdot DW_{0\mu} ]
\\ \label{FLATACTION} \Gamma_{0}[V_0, W_0, A, \eta] &=& g^2T^2 \int
\frac{d\Omega}{4 \pi} X_{0}(Q;V_0, W_0, A, \eta]
\eeq
where
$D_{\lambda}W_{\mu}=\partial_{\lambda}W_{\mu}+ig[A_{\lambda}, W_{\mu}]$~;~
$m^2_g = g^2 T^2 (2N + N_f) / 18 $ and `tr' refers to a trace over the
colour matrices. ~$X_{0}(Q;V_0, W_0, A, \eta]$ is an `angular integrand' of
$\Gamma_{0}$ with $(Q;V_0, W_0, A, \eta ]$ denoting that $X_{0}$ is a
function of $Q$ and a functional of the the various fields.

The Euler-Lagrange equations give
\beq\label{FSEQM}
\funcderv{X_{0}}{V_{0}} = 0   \Rightarrow (Q \cdot D)W_{0\mu}=Q^{\nu}
F_{\mu\nu} \mbox{\hspace {1cm}}
\funcderv{X_{0}}{W_{0}} = 0   \Rightarrow
(Q \cdot D)^2 V_{0\mu}=\smallfrac{3}{2}m_{g}^{2}Q^{\nu}F_{\mu\nu}
\eeq
 We have not yet specified boundary conditions. These are
chosen so that the solutions we obtain from (\ref{FSEQM})  for the auxiliary
fields, when substituted into (\ref{FLATACTION}), yield
the flat space gluon effective action $\Gamma_0^{HTL}[A,\eta]$.
 However, we note that  $\Gamma_0^{HTL}[A,\eta]$ is not uniquely
defined in Minkowski space. [As is well known \cite{taylor}, different
$\Gamma_0^{HTL}[A,\eta]$ correspond to different analytic
continuations of the uniquely defined Euclidean effective action.
 This, in turn, corresponds to a choice of writing non-local
contributions to $\Gamma_0^{HTL}[A,\eta]$ in a retarded or advanced
form]. Thus, the boundary conditions are not uniquely defined by the
requirement
that we derive a flat space gluon effective action.

Suppose we use
homogeneous retarded conditions (i.e. that  $W_{0\mu}~,~V_{0 \mu }
\rightarrow 0 $ as $ t \rightarrow -\infty$). Then we must
take a retarded solution to (\ref{FSEQM}) given by
\beq \label{HJK}
W_{0\mu} = (Q \cdot D)^{-1}_{ret} (Q^{\nu}F_{\mu\nu}) && V_{0\mu} =
 (Q \cdot D)^{-2}_{ret} (Q^{\nu}F_{\mu\nu})
\eeq
where $(Q \cdot D)^{-1}_{ret}$ is an integral operator whose action on an
arbitrary field $G^A$ is  defined by
\beq\label{INVERSE}
 ((Q \cdot D)^{-1}_{ret} G)^A (x) &=&   \int_{- \infty}^0 d\theta
 U^{AB}(x, x+Q \theta ) G^B(x+Q \theta)
 \\ \nonumber     \mbox{with} \qquad U(x, x+Q \theta ) ^{AB} &=& \Big(
     \mbox{P}      \exp  \int_{\theta}^0   d \theta '
                  [-ig Q^\mu{\cal A}_{\mu}(x+Q \theta ')] \Big) ^{AB}
\eeq
The resulting expression on substituting (\ref{HJK}) into
(\ref{FLATACTION}) gives a retarded form of the flatspace gluon
effective action
\beq \nonumber
g^2T^2X^{HTL}_{0}(Q;A, \eta] &=& \int
 d^{4}x 2 \mbox{tr} [-
\smallfrac{3}{4}m_{g}^{2} ((Q \cdot D)^{-1}_{ret} Q^{\lambda}
F_{\mu\lambda})((Q \cdot D)^{-1}_{ret} Q^{\lambda}F^{\mu}_{~~\lambda})  ]
 \\ \label{FSHT}
\Gamma^{HTL}_{0}[A, \eta] &=& g^2T^2 \int
 \frac{d\Omega}{4 \pi} X_{0}(Q;A, \eta]
\eeq

Alternatively using homogeneous advanced conditions
 results in a different, advanced,  expression for
$\Gamma_0^{HTL}[A,\eta]$ above, where the integral operator $(Q \cdot
D)^{-1}_{ret}$ is replaced in (\ref{FSHT}) by $(Q \cdot
D)^{-1}_{adv}$ whose action on an
arbitrary field $G^A$ is  defined by
\beq\label{INVERSEAD}
 ((Q \cdot D)^{-1}_{adv} G)^A (x) &=&  - \int^{ \infty}_0 d\theta
 U^{AB}(x, x+Q \theta ) G^B(x+Q \theta)
\eeq
No other boundary condition  results in a solution of (\ref{FSEQM}),
which when substituted into (\ref{FLATACTION}), yields an expression of
the form of a possible $\Gamma_0^{HTL}[A,\eta]$.


 We impose a constraint on the gluon fields
\beq \label{CN}
        (Q \cdot D)^{-1}_{ret}(Q^\nu F_{\mu \nu}) = (Q \cdot
D)^{-1}_{adv}(Q^\nu F_{\mu \nu}) \equiv (Q \cdot D)^{-1}(Q^\nu F_{\mu \nu})
\eeq
We denote the set of gluon fields which satisfy (\ref{CN}) by ${\cal
R}_0^{gluon}$. For  gluon fields in ${\cal R}_0^{gluon}$ the flat space
 gluon hard thermal loop effective
action
is uniquely  defined and  the initial choice between advanced
and retarded boundary conditions for solutions of (\ref{FSEQM}) above
becomes redundant.

In the following sections we extend the theory to   non-zero
$\kappa$ and encounter several equations  for which
we have to choose between advanced and retarded  boundary conditions,
as  for (\ref{FSEQM})
above. The different choices lead, for arbitrary fields, to effective actions
 corresponding to  different analytic continuations of the uniquely
defined Euclidean effective action. For simplicity, we impose
constraints on the fields such that, as for solutions of (\ref{FSEQM})
 above, any choice between
retarded or advanced non-local functionals of the fields
 is actually redundant.

These constraints imply  that whenever in the following sections  we
encounter a non-local  expression
of the form $\int d\theta' B(x+Q \theta')$, where B  is some
function of the fields $(\bar{\psi},\psi,A,e)$, the fields are
sufficiently constrained to ensure that
\beq
 \int_{-\infty}^\theta d\theta' B(x + Q \theta') =
 - \int^{\infty}_\theta d\theta' B(x +Q \theta')
\eeq
Therefore, we effectively restrict the fields so that in the following
sections we can always take
\beq \label{INTE}
   \int_{-\infty}^{+\infty} d\theta' =0
\eeq
This imposes a number of constraints on the collection of fields
$(\bar{\psi},\psi,A,e)$ and we denote the set of all such collections
by ${\cal R}$. For example, from above we have that that the gluon
fields are restricted to some subset of ${\cal R}^{gluon}_0$.
The effective actions and energy momentum tensors derived below are
 initially valid only for fields in
${\cal R}$ and are uniquely defined on ${\cal R}$.
For fields not in ${\cal R}$, the  effective actions and energy momentum
tensors
can be derived by analytic continuation of the expressions valid in
${\cal R}$.

The collections of fields in ${\cal R}$ satisfy a great number of
constraints, which are outlined below. In the following sections
${\cal R}$ is assumed to be non-empty, unless explicity stated otherwise.

[If ${\cal R}$ is an empty
set, the results of this paper are still valid. The calculations, as
presented below, are invalidated. However, they can easily be repeated
but instead of assuming the fields are in ${\cal R}$, we choose some
prescription  for writing  either an advanced or retarded form for any
non-local expression encountered below (like solutions of
(\ref{FSEQM}) above).
 Different prescriptions  lead to effective
actions corresponding to  different analytic continuations of the uniquely
defined Euclidean effective action. The effective action and QGP
energy momentum tensor derived using this method are exactly the same
as a specific analytic continuation of the effective action and QGP
energy momentum tensor derived below. As will be verified below, there
is a prescription
which yields an effective action  for fields not in ${\cal
R}$ with analyticity properties such that its associated energy
momentum tensor is retarded. (The physical significance of retarded
currents for the QGP plasma is discussed by Jackiw and Nair
\cite{jackiw})].


\subsection{Inclusion of weakly coupling graviton field}

Consider a generalisation of the above Minkowskian theory  to include
interactions of the QGP with  a weak graviton
field, $\varphi ^{\alpha \beta}$.
 Following \cite{taylor(b)} and \cite{taylor} we
assume the spacetime to be asymptotically Minkowskian.
By considering the symmetries of this theory, we aim to write down an action
of local auxiliary fields which, on
elimination of the  auxiliary fields, gives  to $O(\kappa)$ the generating
functional \ghtl
of hard thermal loop gluon-graviton interactions, i.e.
\beq
\lefteqn{\ghtl [A, g]=\int \frac{d\Omega}{4 \pi} \sum_{m, n}
d^4x_{1}, \ldots, d^4x_{m}d^4y_{1}, \ldots, d^4y_{n} } \\ \nonumber
&& \hspace{-1.0cm}A_{\mu_1}(x_{1}), \ldots, A_{\mu_m}(x_{m}),
 \varphi^{\alpha_1
\beta_1}(y_{1}), \ldots, \varphi^{\alpha_n \beta_n}(y_{n})
G^{HTL}(x_{1}, \ldots, x_{n}, y_{1}, \ldots, y_{n};Q)^{\mu_1 \ldots
\mu_m}_{\alpha_1 \beta_1 \ldots \alpha_n \beta_n} \\  & \equiv &
g^2T^2 \int
\frac{d\Omega}{4 \pi} X^{HTL}(Q;A, g]
\eeq
  $G^{HTL}$ refers to the hard thermal 1-loop truncated diagram with $n$
external  graviton fields and $m$ external gluon fields. As  $\kappa\ll1$
we  take $n=1$ unless explicity specified otherwise.
Clearly, the definition of $X^{HTL}$ above only specifies $X^{HTL}$ up
to a total angular differential. However it is known from hard thermal
loop field loop theory \cite{taylor}  that $X^{HTL}$ can be chosen to
have the following properties
\
\begin{enumerate}
      \item[A. ] It is non-local but with non-localities only of the form of
 products of $(Q \cdot
\partial)^{-1}$ operators. Examples of such operators are given by
  (\ref{ADRE})
      \item[B. ] It is homogeneous of degree zero in Q.
      \item[C. ] It has dimensions of (energy)$^{-2}$.
      \item[D. ] It is invariant under general coordinate, SU(N) gauge ,  Weyl
scaling (and for spinor theories) local Lorentz transformations which
are restricted to tend to the identity at infinity. (A property which
all local transformations in this paper are assumed to obey).

\end{enumerate}
 When  $X^{HTL}$ is referred to in the rest of this paper we implicitly mean
 the
form of $X^{HTL}$ which satisfies the above conditions. [An alternative
expression differing by a total angular differential is the basis of
the application of Chern-Simons field theory to hard thermal loop
phenomena \cite{nair}].
In an attempt to produce $X^{HTL}$ at least to  $O(\kappa)$ using an
auxiliary field lagrangian method, thus enabling a simpler calculation of
$T^{HTL}_{\alpha \beta }$,  we extend
(\ref{FLATACTION}) to form a general coordinate and Weyl scaling
invariant expression.
However, we find that is not possible to perform this extension uniquely and
 that this
non-uniqueness has interesting implications, which will be elaborated below.

\subsubsection{ A Weyl scaling and general coordinate extension}

Consider the general coordinate and Weyl scaling invariant extension :
\beq
\label{EACT} \nonumber
g^2T^2X_{E}(Q;V, W, A, g] &\equiv& \int  \meas  2
\mbox{tr} \Bigg[ -\WW  \\ &+& V_{\mu}\dot{y}^{\lambda}F_{\nu\lambda}
 -V_{\nu}[\YDW] \\ \nonumber && \hspace{-1.8cm}+E \Big\{ V_{\nu} [\ambig]
\Big\}
\Bigg] \\ \nonumber \Gamma_{E}[V, W, A, g] &\equiv & g^2T^2
\int \frac{d\Omega}{4 \pi} X_{E}(Q; V, W, A, g]
\eeq

where $E$ is an arbitrary constant and $\nabla_{\lambda}$ refers to a
space-time and colour adjoint  derivative i.e.
\beq \nonumber
\nabla_{\lambda}W_{\mu} = \partial_{\lambda}W_{\mu} - \Gamma_{\lambda
\mu}^{\rho}W_{\rho}  +ig[A_{\lambda}, W_{\mu}]
\eeq
Note that $\nabla$ acting on a colour trivial object, as will occur
below,  reduces to the
standard space time derivative.


In order to define $\dot{y}^\mu$ above, we must consider the null
geodesic, denoted by $y^\mu (x,\theta)$, which passes through $x$, is
affinely parameterised by $\theta$
and satisfies the following conditions
\beq \label{NGC}
y^\mu(x,0) = x^\mu && \dot{y}^\mu(x,\theta) \equiv
\frac{d y^\mu (x,\theta)}{d \theta}
\rightarrow Q^\mu \mbox{~~ as  ~~} \theta \rightarrow \pm \infty
\eeq
The tangent vector of $y^\mu (x,\theta)$ at $\theta = 0$ defines the
expression $\dot{y}^\mu$ used
in (\ref{EACT}) i.e.
\beq
 \dot{y}^\mu \equiv \dot{y}^\mu(x,\theta ) \mid _{\theta =0} \equiv
\frac{d y^\mu (x,\theta)}{d \theta} \Big| _{\theta =0}
\eeq
  Note that $\dot{y}^\mu$  generalises  $Q^\mu$.
Using the null geodesic equation
\beq \label{NGE}
\ddot{y} (x,\theta)   = - \Gamma^{\lambda}_{\mu \nu}
 (y(x,\theta))\dot{y}^{\mu}(x,\theta) \dot {y}^{\nu}(x,\theta)
\eeq
and the conditions (\ref{NGC}) we can expand $\dot{y}^\mu(x,\theta)$ in powers
in of $\kappa$. Using (\ref{INTE}), we have to first order in $\kappa$
\beq \label{YDOTT}
\dot{y}^{\lambda}(x,\theta) &=& Q^{\lambda} - \kappa \int_{\theta}^{\infty}
 d\theta '
\gamma^{\lambda} ( x + Q\theta ') +O(\kappa^2) \\ \nonumber &=&
Q^{\lambda} + \kappa \int^{\theta} _{-\infty} d\theta '
\gamma^{\lambda} ( x + Q\theta ') +O(\kappa^2) \\ \nonumber
 \mbox{where } \spp \gamma^{\lambda} &=&
-Q_{\mu} ( Q \cdot \partial ) \phi^{\lambda \mu} + \smallfrac{1}{2}
Q_{\mu} Q_{\nu} \partial^{\lambda} \phi^{\mu \nu}
\eeq
Note that $\int^{\theta} _{-\infty} d\theta '
\gamma^{\lambda} ( x + Q\theta ') = - \int_{\theta}^{\infty} d\theta '
\gamma^{\lambda} ( x + Q\theta ')$ gives  a constraint on the vierbien
associated with the metric.  This constraint therefore must be
satisfied by   any vierbien belonging to
${\cal R}$. Similar restrictions arise at higher orders in
$\kappa$. Without such restrictions we can choose $\dot{y}^\lambda(x,\theta)$
such that $(\dot{y}^\lambda (x,\theta) - Q^\lambda)$ satisfies either
homogeneous retarded or homogeneous advanced boundary conditions, but
not both.

$\Lambda$ is a dimensionless scalar, homogeneous of degree zero in Q, which
 vanishes when $g = \eta$  with the
Weyl scaling property that $\Lambda\rightarrow \Lambda + 2 \sigma$.

[Recall that a Weyl scaling is given by $g^{\alpha \beta} \rightarrow
e^{2 \sigma}  g^{\alpha \beta} $ , $  g_{\alpha \beta} \rightarrow
e^{-2 \sigma} g_{\alpha \beta} $ , $ A_{\mu} \rightarrow A_{\mu}$ , $
\psi \rightarrow e^{ \frac{3\sigma}{2} } \psi$ where for
this paper we assume $ \sigma \rightarrow 0$ as $x \rightarrow \infty$].

We can give a simple closed form expression for $\Lambda$ in terms of
$\dot{y}^\mu (x,\theta )$ defined above. First recall  that  a null
geodesic, although  invariant under  a Weyl scaling, is reparameterised so
 that its affine
parameter $\theta$ transforms as $d\theta \rightarrow e^{-2 \sigma}
d\theta$. Thus
\beq  \label{YWS}
      \dot{y}^\mu(x,\theta) \equiv \frac{d y^\mu(x,\theta)}{d \theta}
\rightarrow e^{2 \sigma(y(x,\theta))} \dot{y}^\mu (x,\theta)
\eeq
on a Weyl scaling .

Taking into account the well known (see eg. \cite{hawk}) Weyl scaling
properties of the
Christoffel symbol, it is straightforward to show that
\beq
      (\nabla_\alpha \dot{y}^\alpha) \rightarrow e^{2 \sigma} (
\nabla_\alpha \dot{y}^\alpha -2(\dot{y}^\alpha \partial _ \alpha
\sigma ))
\eeq
on Weyl scaling and hence $\Lambda $ can be defined by the solution of
the equation
\beq
    (\dot{y} \cdot \partial) \Lambda = - (\nabla_\alpha \dot{y}^\alpha)
\eeq
Again using (\ref{INTE}), we have
\beq \label{LAM}
    \Lambda(y(x,\theta))  = + \int_\theta ^{\infty} d \theta '
 ~\nabla_{\nu}\dot{y}^{\nu}(y(x,\theta ')) =
 - \int^\theta _{-\infty} d \theta '
{}~\nabla_{\nu}\dot{y}^{\nu}(y(x,\theta '))
\eeq
 Again the equality of advanced and retarded versions of $\Lambda$
requires a constraint on the vierbien fields belonging to ${\cal R}$.
 We have
\beq \label{LWS}
\Lambda(x) \equiv \Lambda (y(x,0)) \rightarrow \Lambda (y(x,0 )) + 2 \sigma
(x) = \Lambda (x) + 2 \sigma (x)
\eeq
 on Weyl scaling as required. It
is also straightforward to check this  definition is consistent
with all the other properties required of $\Lambda$.


Note however $\Lambda$ is not uniquely defined at $O(\kappa^{2})$ and
the consequences of this are discussed in Appendix C. Below, we use
$\Lambda$ as given in (\ref{LAM}). However, as we are working at
$O(\kappa)$ throughout this paper (except briefly in Appendix C), all
of the following discussion is independent of the second order
ambiguity in $\Lambda$.

$X_{E}(Q; V, W, A, g]$ above has been constructed to be Weyl scaling
invariant if $V$ and $W$ are Weyl scaling invariant.  [Note that
although the Weyl scaling properties imposed on W and V appear to be
 arbitrary , this apparent freedom doesn't
actually affect the Weyl scaling and
general coordinate invariant extension of the flat space action at
$O(\kappa)$, as demonstrated in Appendix B].
\newline
To insure Weyl scaling invariance of $X_{E}$ it is not
sufficient to generalise $ Q \cdot DW_{\mu}$
to $ \dot{y} \cdot \nabla W_{\mu}$
due to the awkward Weyl scaling properties of the Christoffel symbols
present in $\nabla$ and hence extra terms are added so that
\beq \nonumber
 Q \cdot DW_{\mu} \rightarrow \YDW = \dot{y}^{\alpha}\partial_{\alpha}W_{\mu}
  +  W_{\nu}\partial_{\mu}\dot{y}^{\nu} - \dot{y}^{\alpha}W_{\alpha}
\Lambda, _{\mu}
\eeq
Thus the Christoffel symbol of $W_{\nu} \nabla_{\mu} \dot{y}^{\nu}$ cancels
the Christoffel
symbol of $\dot{y}^{\alpha} \nabla_{\alpha} W_{\mu}$ in a covariant
manner. Using (\ref{YWS}) and and (\ref{LWS}) is is straightforward to
verify that the above expression is simply
multiplied by $e^{2\sigma}$ on Weyl scaling.
This ad-hoc fixing of  Weyl scaling is not unique and an alternative
is given by
\beq \nonumber
& Q \cdot DW_{\mu} = \eta_{\mu \nu} Q \cdot DW^{\nu} \rightarrow g_{\mu\nu}
[\dot{y}^{\alpha}\nabla_{\alpha} W ^{\nu}  -
W^{\mu}\nabla_{\mu}\dot{y}^{\nu} - W^{\nu} \dot{y} \cdot \partial
\Lambda + \dot{y}^{\nu}W \cdot \partial \Lambda ] &
\eeq

The difference between these two possibilities gives the coefficient
of E in (\ref{EACT}) above and thus E is an arbitrary parameter,  reflecting
the ambiguity present in extending Weldon's action. Of course,
we don't know yet for what value of E (if any) (\ref{EACT}) will yield
$X^{HTL}$ (to $O(\kappa)$ at least) on elimination of the auxiliary
fields. Further ambiguities exist at $O(\kappa^2)$ and are discussed
in Appendix C.

\subsubsection{Elimination of the auxiliary fields}

Let $X_{E}(Q;A, g]$ and $\Gamma_{E}[A, g]$ denote the angular integrand
and the action formed by the elimination of the auxiliary fields from
$X_{E}(Q;V, W, A, g]$ and $\Gamma_{E}[V, W, A, g]$ using the Euler-Lagrange
 equations. Thus,
\beq \label{XW}
g^2T^2X_{E}(Q;A, g] &=&  \int \meas 2\mbox{tr} (-\WW) \\ \nonumber
\Gamma_{E}[A, g] &=& g^2T^2 \int \frac{d\Omega}{4 \pi} X_{E}(Q;A, g]
\eeq
    where W is a solution of the Euler Lagrange equations
\beq\label{EQMW}
    \dot{y}^\alpha \nabla_\alpha W_\mu &=& \dot{y}^\nu F_{\mu \nu} +
W_\alpha (\Delta ^E)^\alpha_\mu \\ \nonumber
  (\Delta ^E)^\alpha_{~~\mu} &=& \dot{y}^\alpha \Lambda ,
_{\mu} - \nabla_\mu \dot{y}^\alpha + E [\nabla_\mu \dot{y}^\alpha
-\dot{y}^\alpha \Lambda ,
_{\mu} + \smallfrac{1}{2} g_\mu^\alpha \dot{y} \cdot \partial \Lambda
+ (\alpha \leftrightarrow \mu) ]
\eeq

Clearly any solution $W_\mu$ to the above equation is covariant under
general coordinate transformations, transforms like the field strength
under gauge transformations and has dimensions L$^{-1}$. A solution of
(\ref{EQMW}) can be obtained as a perturbative expansion in
$\kappa$. Writing
\beq
W_\mu = \sum\limits_{n=0}
\kappa^n W_{n \mu}~, & (\dot{y} \cdot \nabla) = Q
\cdot D + \sum\limits_{n=1} \kappa ^n
(\dot{y} \cdot \nabla)_n~, & \dot{y}^\nu
 = Q^\nu + \sum\limits_{n=1} \kappa ^n \dot{y}^\nu_n
\\ \nonumber & (\Delta ^E)^\alpha
_\mu = \sum\limits_{n=1} \kappa^n (\Delta ^E _n)^\alpha _\mu
\eeq
we have
\beq \label{EQMWW}
W_{0\mu} &=& (Q \cdot D)^{-1} (Q^\nu F_{\mu \nu}) \\ \label{EQW}
W_{n\mu} &=& (Q \cdot D)^{-1}  \{ \dot{y}_n^\nu F_{\mu \nu}
+ \sum_{p=0}^{n-1} [- (\dot{y} \cdot
\nabla)_{n-p} W_{p\mu} + (\Delta ^E _{n-p})^\alpha _\mu W_{p \alpha}] \}
\eeq
 Thus $W_{0\mu}$ agrees with the flat space auxiliary field
 in section (2.1). We impose sufficient restrictions on the gluon and
vierbien fields in  ${\cal R}$ to ensure that  the choice of a retarded
 or advanced inverse
in (\ref{EQW}) is redundant.

By expanding $(Q \cdot D)^{-1}$ in powers of $g$, it is straightforward
to verify that each
$W_{n\mu}$ and hence $W_\mu$ consists of non-localities of the form
 of products of $(Q \cdot \partial)^{-1}$. Applying
$Q\cdot\frac{\partial}{\partial Q}$ to both sides of equation
(\ref{EQW}) we see (by inducting on $n$) that each $W_{n\mu}$ is
homogeneous of degree zero in $Q$ and hence so is $W_\mu$.

$W$ is also Weyl invariant as this is imposed on $W$ when making the
extension from flat space. Note that any solution of (\ref{EQW}) must
be Weyl invariant as it can be shown, by induction on $n$, that each
$W_{n \mu}$ is Weyl invariant.

Hence $W_{\mu}$ has the correct properties to ensure $X_{E}(Q;A, g]$
obeys (A) $\rightarrow$ (D) as given in section (2.2).
This fact, together with the observation that in the limit
$g\rightarrow \eta , ~X_{E}$ reduces to $X_{0}$,
is interesting and contrary to the  previous assertion  that a functional
satisfying conditions (A) $\rightarrow (D)$ and  with the
correct flatspace terms would be in fact equal to $ X^{HTL}(Q;A, g]$ at
higher orders in $\kappa$ \cite{taylor,taylor(a)}. This is
considered in more detail in section (A.2) of Appendix A.

\subsubsection{The relation between $X_E(Q;A,g]$ and $X^{HTL}(Q;A,g]$}

We show in Appendix A that the following holds: given any
expression  which

\begin{enumerate}
     \item[i. ]   Satisfies conditions (A) $\rightarrow$ (D) above.
     \item[ii. ]  Reduces to $X^{HTL}_{0}(Q;A, \eta]$ when $g\equiv \eta$.
     \item[iii. ] Agrees with $X^{HTL}(Q;A, g]$ \quad at \quad $O(g^{2}
\kappa )$.
\end{enumerate}
\noindent
must, at least to $O(\kappa)$, be equal to $X^{HTL}(Q;A, g]$  for all orders
 in $g$.

Now consider  $X_{E}[A, g]$ above.  We know it satisfies conditions
(i) and  (ii) above. We will
show that for $E=0$ ,  it also agrees with $X^{HTL}$ at $O( g^{2} \kappa )$.
 First, we calculate  $S^{E}_{\alpha \beta}$ defined by
\beq \nonumber
& S_{\alpha\beta}^{E} \equiv ~2g^2T^2 \mbox{lim} _{g\rightarrow \eta}
 \left(\funcderv{X_{E}(Q;A, g]}{g^{\alpha\beta}} \right)
\eeq
(so that the angular integral of the above is equal to the energy
momentum tensor associated with the action $\Gamma_{E}$). Then we
 explicitly verify that for {\bf $E=0$ }, $S_{\alpha
\beta}^{E=0}$ is equal to the $O(g^2)$ contribution to the angular
integrand of $T^{HTL}_{\alpha \beta }$ which is
consistent with conditions (A) $\rightarrow$ (D) (as
calculated using
 thermal field theory in  \cite{taylor}). This shows equality between
$X_{E}$ and $X^{HTL}$ at $O(g^2
\kappa)$ which explicitly confirms condition (iii). Hence we deduce that
  $X_{E=0}(Q;A, g] = X^{HTL}(Q;A,g] $ at
least to $O(\kappa)$ and hence  that
 $\int d\Omega (4 \pi ) ^{-1} ~ S ^{E=0}_{\alpha \beta}$ is equal to
 the gluon contribution of the retarded QGP energy momentum tensor,
$T^{HTL}_{\alpha \beta }$, to all orders in $g$.

\subsection{Calculation of $S_{\alpha \beta}^{E}$ and the QGP energy
momentum tensor}
The calculation of  $S^{E}_{\alpha \beta}$
 is simplified by noting that for when $V$, $W$ are considered as
functionals of the physical fields ie.  $V=V(Q;A, g]$, $W=W(Q;A, g]$ we have
\begin{eqnarray*} \textstyle
\left.  \funcderv{X_{E} (Q;V, W, A, g]}{g^{\alpha\beta}} \right|_{A, V, W}
&=& \textstyle
\left.  \funcderv{X_{E} (Q;V, W, A, g]}{g^{\alpha\beta}} \right|_{A}
       - \left. \funcderv{X_{E}(Q;V, W, A, g]}{V_{\mu}}
\right|_{A, g, W}\funcderv{V_{\mu}(Q;A, g]}{g^{\alpha\beta}}  \\ &-&
\textstyle \left. \funcderv{X_{E}(Q;V, W, A, g]}{W_{\mu}}  \right|_{A, g, V}
\funcderv{W_{\mu}(Q;A, g]}{g^{\alpha\beta}}
\end{eqnarray*}

\noindent but with  $g=\eta$,  $V=V_{0}$,  and $W=W_{0}$ (see
(\ref{FSEQM}))   we have $\funcderv{X_{E}(Q;V, W, A, g]}{V_{\mu}}$ and
$\funcderv{X_{E}(Q;V, W, A, g]}{W_{\mu}}$ zero and hence
\beq\label{EMT2}
S_{\alpha\beta}^{E} = 2g^2T^2 \mbox{lim}_{g \rightarrow \eta} \Big(
\left.   \funcderv{X_{E} (Q;V=V_{0}[A, g], W=W_{0}[A, g], A, g]}
{g^{\alpha\beta}} \right|_{A, V, W const} \Big)
\eeq
Note that calculating $S_{\alpha\beta}^{E}$ only requires the flat
space equations for the auxiliary  fields and thus that $X_{E}(Q;A,g]$
at $O(\kappa)$ depends only on
the flat space auxiliary fields. Performing the
calculation it is useful to note the following
\beq \label{ADRE}
 \int d^4x  G(x)((Q \cdot \partial)^{-1}_{ret} F) (x)  &=& -\int d^4x
((Q \cdot \partial)^{-1}_{adv} G) (x) F(x) \\ \nonumber \mbox{where }
((Q \cdot \partial)^{-1}_{ret} F) (x)  = \int_{- \infty}^0 d\theta
 F(x+Q \theta)
&& ((Q \cdot \partial)^{-1}_{adv} F) (x) =  -\int_0^{\infty} d\theta
 F(x+Q \theta)
\eeq
The labels `$ret$' and `$adv$' can be dropped if which  $\int d\theta'
 F(x+Q \theta')$ and $\int d\theta' G(x+Q \theta')$
are non-local functionals of the fields for which (\ref{INTE})
holds. We impose sufficient restrictions for fields in ${\cal R}$ to
ensure that all non-local functionals encountered below satisfy (\ref{INTE}).

Hence from the definitions of
$\dot{y}$ and $\Lambda$, we have for any $F_\lambda$ and $F$
\beq \label{ADR}
\int dx \dot{y}^{\lambda}F_{\lambda} &=& \int dx \phi^{\alpha \beta}
\smallfrac{1}{2} \EMYDOT F_{\lambda}
\\ \nonumber\int dx \Lambda F &=& \int dx \phi^{\alpha
\beta}\smallfrac{1}{2} \EMETA F   \\  \nonumber \mbox{where }
 A_{ \{ \alpha } B_{ \beta \} } &=& A_{\alpha}B_{\beta} + A_{\beta}B_{\alpha}
\eeq

On performing the calculation we find
\beq
\nonumber
\lefteqn{S_{\alpha\beta}^{E}  = 4\mbox{tr} \Big( \smallfrac{1}{2}
\EMLAM [-\WCW]
-\WAWB  } \\ \label{EMT3} &   &
  +\smallfrac{1}{2} \EMYDOT [V_{0}^\mu F_{\mu\lambda}
- V_{0\mu}D_{\lambda}W_0^\mu + \partial_{\mu} (W_{0\lambda} V_{0}^{\mu})]
\\ \nonumber &   &   +\smallfrac{1}{2}E \Big\{ [ Q\cdot \partial
(W_{0\{ \beta } V_{0 \alpha \} } )- \EMYDOT[\partial_{\mu}(W_{0\lambda}
vV_{0}^{\mu} + V_{0\lambda} W_{0}^{\mu})] \\ \nonumber &  &   - \EMETA
[Q\cdot \partial ( V_{0\mu} W_{0}^{\mu} )] \Big\} \Big)
\eeq
\noindent
Note  $V_{0 \mu}$ and $W_{0 \mu}$ are given by (\ref{FSEQM}) and we
have used $Q^\mu V_{0 \mu} = Q^\mu W_{0 \mu}=0$ in deriving the
above. The
the  coefficient of $E$ in (\ref{EMT3}) is discussed further in
section A.2 of Appendix A.

Setting E=0 we find $S^{E=0}_{\alpha\beta}$ gives equation (6) of
 \cite{taylor} in momentum space at $O(g^{2})$ thus showing,
given the result of Appendix A,  that $X_{E=0}(Q;A, g]=X^{HTL}(Q;A,g]$
at $O(\kappa)$ for all orders in $g$.
Thus $S^{E=0}_{\alpha\beta}$ is the angular integrand of the
 gluon contribution to the
 QGP energy momentum tensor
$T^{HTL}_{\alpha \beta }$ and hence
\beq
T^{HTL}_{\alpha \beta } = \int \frac{d\Omega}{4 \pi} S^{E=0}_{\alpha \beta}
\eeq
The calculations deriving  this are valid for gluon fields in
${\cal R}$. We can analytically continue from ${\cal R}$ to consider
this expression for gluon fields not in ${\cal R}$.
 A choice of analytic continuation where all fields are retarded is of
particular physical significance \cite{jackiw}. Despite using a local
 auxiliary field angular
integrand, $T^{HTL}_{\alpha \beta }$ is not local even with respect to
the auxiliary fields. This is because $\dot{y}^{\mu}$ and $\Lambda$
have a non-local dependence on the graviton fields.

An interesting but difficult question is whether or not $X_{E=0}(Q;A, g]$
is correct at $O(\kappa^{2})$ and some of the difficulties that arise
are discussed in Appendix C.

[We again consider the problem of choosing a prescription which yields
a retarded energy momentum tensor. As discussed in section (2.1),
such an approach is required if we drop  the assumption that ${\cal
R}$ is non-empty. The prescription is to choose the advanced expressions
for
$\dot{y}$ and $\Lambda$ in (\ref{YDOTT}) and (\ref{LAM}), while  choosing
retarded expressions for all other possibilities.
With this choice of prescription we see that, on use of (\ref{ADRE}),
(\ref{ADR}) is still valid on replacing $(Q \cdot \partial)^{-1}$ by
$(Q \cdot \partial )^{-1}_{ret}$. This would therefore lead to
retarded energy momentum tensor, automatically valid for fields not in
${\cal R}$].

\section{Single auxiliary field model}
\subsection{Flat space auxiliary fields}
The same analysis is applied to a model with only one auxiliary field
$m_{0}^{\mu}(Q, x) = m_{0}^{\mu A}(Q, x) t^A$ which again transforms like the
field strength on gauge transformations.

Consider the following angular integrand $X_{0}(Q;m_0, A, \eta ]$ and
action $\Gamma_{0}[m_0, A, \eta ]$, as flat space precursors:
\beq
 g^2T^2X_{0}(Q;m_0, A, \eta] &=& \int d^{4}x (4b) \mbox{tr}
[m_{0}^{\lambda}Q^{\nu}F_{\lambda\nu} + \smallfrac{1}{2}
(Q\cdot Dm_{0 \nu})(Q\cdot Dm_0^{\nu})]
\\ \nonumber \Gamma_{0}[m_0, A, \eta]  &\equiv&  g^2T^2
\int \frac{d\Omega}{4 \pi} X_{0}(Q; m_0, A, \eta]
\eeq
\noindent
 On taking $b=\frac{3}{4}m^{2}_{g}$ and using  (\ref{INTE})
together with  the
Euler-Lagrange equation  for $m$, the
above reduce to the non-local flat space
hard thermal loop angular integrand and action, i.e.
\beq \label{FSMEQM}
 & \funcderv{X}{m} =0 \quad \Rightarrow \quad m_{0\mu}=
(Q\cdot D)^{-2}Q^{\nu}F_{\mu\nu} \quad \Rightarrow
  &\\ \nonumber & g^2T^2X^{HTL}_{0}(Q;A, \eta]   = -\int d^{4}x ~~
2 \mbox{tr}[\smallfrac{3}{4}m_{g}^{2} [(Q \cdot D)^{-1}
 Q^{\lambda}F_{\mu\lambda}][
(Q \cdot D)^{-1} Q^{\lambda} F^{\mu}_{~~\lambda} ]]
\\ \nonumber & \Gamma_{0}[A, \eta]  \equiv  g^2T^2 \int
\frac{d\Omega}{4 \pi} X_{0}(Q; A, \eta] &
\eeq
 Note
that above we have used
\beq
\int d^4x (Q^\nu F^\mu _{~~ \nu})[(Q \cdot D)^{-2} (Q^\nu F_{\mu \nu})] =
-\int d^4x  [(Q \cdot D)^{-1}(Q^\nu F_{\mu \nu})][(Q \cdot D)^{-1}(Q^\nu
F^\mu _{~~\nu})]
\eeq
which is valid for fields in  ${\cal R}$.

\subsection{Inclusion of Weakly coupling graviton field}
Again, on generalising to curved space and imposing Weyl
invariance , we hope to obtain $X^{HTL}[A, g]$ on the elimination of the
auxiliary fields.  As in section 2 we find there are
ambiguities.  Consider the generalisation where $E'$ is an arbitrary constant:
\beq \label{MACT}
 g^2T^2X_{E'} &=& \int \measM 2 \mbox{tr} \Big[ 2b (\YFETAM)
 \\ \nonumber & & \hspace{2.0cm} + b~g_{\mu \nu}(B^\mu B^\nu + E'
B^\mu C^\nu + \smallfrac{1}{2} E'^{2}      C^\mu C^\nu )\Big]
 \\ \label{BEQN}  \mbox{where } B^\mu &=& \YDMM \\ \label{CEQN}
 C^\nu &=& \AMBIGUITYM
\eeq

\noindent
with $m^{\mu} \rightarrow m^{\mu}$ on Weyl scaling. Again the choice of
the Weyl scaling properties doesn't affect $X_{E'}(Q;m, A, g]$ at
$O(\kappa)$ (see Appendix B) and the choice above is for simplicity.

The coefficients of $E'$ \& $E'^2$ correspond to to the fact that the
extension to a general coordinate and Weyl scaling invariant action is
not unique and they are  due to  taking  different combinations
of the following two
possibilities for the generalisation of $ Q \cdot D m^{\mu} $ (both of
which are simply  multiplied by $e^{2 \sigma}$ on Weyl scaling).
\beq \nonumber
  Q \cdot D m^{\mu} &\rightarrow& \dot{y} ^\alpha \nabla_\alpha
m^{\mu} - m^\alpha
\nabla_\alpha  \dot{y}^{\mu} + \dot{y}^{\mu} m^\alpha
\partial_\alpha \Lambda   \\
\nonumber
  Q \cdot D m^{\mu} = \eta^{\mu\nu} Q \cdot D m_{\nu}  &\rightarrow&
g^{\mu\nu} [\dot{y} ^\alpha \nabla _\alpha m_{\nu} + m_{\rho} \nabla_{\nu}
\dot{y}^{\rho} - m^\alpha \dot{y}_\alpha \Lambda, _{\nu} + m_{\nu}
\dot{y} ^\alpha \partial_\alpha \Lambda ]
\eeq

 The
Euler-Lagrange equations for $m^\mu$ are
\begin{eqnarray} \label{MEQM}
\dot{y} \cdot \nabla (B_{\alpha}+ E' C_\alpha) &=& \dot{y}^\lambda
F_{\alpha \lambda} +
(B_{\nu}+ E' C_\nu)  (\Delta ^{E'})^\alpha_{~~\mu}
\end{eqnarray}
\noindent
where $(\Delta ^{E'})^\alpha_\mu $ is given by (\ref{EQMW}) (for $ E
\rightarrow E'$). Thus by (\ref{EQMW}) we see that $W_\mu$ solves
(\ref{MEQM}) and hence we have
\beq
\dot{y}^\alpha \nabla_\alpha m ^\mu &=& g^{\mu \nu} W_\nu -
                                      (\Delta ^{E'})^\mu _{~~\nu} m^\nu
\eeq
This is solved in exactly the same way as (\ref{EQMW}) (after raising
indices and  replacing $\dot{y}^\nu F^{\mu}_{~~ \nu}$ by $g^{\mu \nu
}W_\nu $). Clearly any solution $m^\mu$ to the above equation is
covariant under
general coordinate transformations, transforms like the field strength
under gauge transformations and has dimensions L$^{0}$. By expanding
in powers of $g$ and $\kappa$ it is straightforward to show that it
non-localities are only of the form of products of $(Q \cdot
\partial)^{-1}$. Writing $m^\mu = \Sigma_{n=0} \kappa^n m_n^\mu $,it
is also straightforward to prove [by induction on n exactly as in
section (2.2.2)] that $m^\mu$ is homogeneous in $Q$ of degree minus one
and that its imposed Weyl invariance is consistent with the Euler
Lagrange equations.

 Defining $X^m_{E'}(Q;A,g]$  by (\ref{MACT}) {\em at the solution of
(\ref{MEQM})} we see that $m^\mu$ has the correct properties to ensure
that $X^m_{E'}(Q;A,g]$ satisfies condition (i). Section (3.1) shows
that it satisfies
condition (ii). Thus we see that $X^m_{E'}(Q;A,g]$ is a counterexample
to the assertion  that a functional
satisfying conditions (A) $\rightarrow (D)$ and  with the
correct flatspace terms would be in fact equal to $ X^{HTL}(Q;A, g]$ at
higher orders in $\kappa$ \cite{taylor,taylor(a)}. [In fact, as we will
show below $X^m_{E'}(Q;A,g]$ = $X_{E}(Q;A, g]$ at $O(\kappa)$ for
$E=E'$].

 In addition condition (iii) is satisfied for $E'=0$. We show this by
simply calculating
\beq
      ^mS^{E'}_{\alpha \beta} &\equiv& 2g^2T^2 \mbox{ lim}_{g
\rightarrow \eta }  \funcderv{X^m_{E'}(Q;A,g]}{g^{\alpha \beta}}
\eeq
 and  verifying that $^mS^{E'=0}_{\alpha
\beta} = S^{E=0}_{\alpha \beta}$ (which is equal to the angular
integrand of $T^{HTL}_{\alpha \beta }$
which satisfies (A)  $\rightarrow$
(D) by the results of section 2).

Using the techniques of section (2.3) we find $^mS^{E'}_{\alpha\beta}$
is a function of the gluon fields given by
\beq \nonumber
^mS^{E'}_{\alpha\beta} \! &=& \! 4b \mbox{tr}  \Big\{ \smallfrac{1}{2}
\EMLAM [m_{0\mu}f^{\mu} + \smallfrac{1}{2}
Q\cdot Dm_{0\mu}Q\cdot Dm_{0}^{\mu}]   \\  \label{MTEN}
              &+& \smallfrac{1}{2} \EMYDOT
[m_{0\mu}F^{\mu}_{~~\lambda}+ (D_{\lambda}m_{0\mu})Q\cdot Dm_{0}^{\mu}
\\ \nonumber && \hspace{5.0cm} + \partial^{\mu} (m_{0\mu}
Q\cdot Dm_{0\lambda})]
        -  \smallfrac{1}{2} (Q\cdot D m_{0\alpha}Q\cdot Dm_{0 \beta})
\\ \nonumber &+& \smallfrac{1}{2} E' \Big[
- \EMYDOT[\partial_{\mu}(Q \cdot m_{0\lambda}
m_{0}^{\mu} + m_{0\lambda} Q \cdot m_{0}^{\mu})] \\ \nonumber &-&  \EMETA
[Q\cdot \partial ( m_{0\mu} Q \cdot m_{0}^{\mu} )] \\ \nonumber &+&
Q \cdot \partial (Q \cdot m_{0\{ \beta } m_{0 \alpha \} } )\Big]
 \Big\}
\eeq\noindent
where $f^\mu = Q^\rho F^{\mu} ~ _\rho$ and $m_{0 \mu}$ is given by
(\ref{MEQM}). Note that we have used
$Q^\mu m_{0 \mu}=0$ in deriving the above and that the
 coefficient of $E'$ in $^mS^{E'}_{\alpha \beta}$ above
is the same  as the coefficient of $E$ in (\ref{EMT3}) which  is
consistent with uniqueness results stated in section (2) (and derived
in Appendix A).

Setting $E'=0$ in
(\ref{MTEN}), applying the identity
\begin{eqnarray*}
 (D_{\lambda} m_{0 \mu}) Q \cdot D m_{0 \mu}
 &=& \partial_{\lambda} \big\{ (Q\cdot \partial)^{-1}
 (Q \cdot \partial) [(Q \cdot D m_0 ^\mu)m_{0
\mu}] \big\} - m_0 ^\mu D_{\lambda}(Q \cdot D m_{0 \mu})
\end{eqnarray*}
in the second line above and noting that $W_{0 \mu}=(Q \cdot D)m_{0
\mu}$ and $V_{0 \mu} = 2b~Q \cdot Dm_{0 \mu}$  it is straightforward to see
that $^m S^{E'=0}_{\alpha \beta}$ is equal to $S^{E=0}_{\alpha\beta}$
derived earlier.

 Hence $X^m_{E'=0}(Q; A, g]$ satisfies conditions (i) $\rightarrow$
(iii) . Indeed, we see from the equivalence of $S^E_{\alpha \beta}$ and
$^mS^{E'}_{\alpha \beta}$ (for $E=E'$) that the actions  presented  in
 section (2)
and this section are equivalent at $O(\kappa)$ and hence that  the
results of section (2) could be derived, ab initio, with  a model using
only one auxiliary field.

Again,  considering agreement of these actions and the gluon graviton
 hard thermal loop
effective action at $O(\kappa^{2})$ is a more difficult problem.


\section{ Quark Sector}
\subsection{Flat space auxiliary fields}
Again we perform the analysis of section 2 only now for the quark
sector of the theory, with the aim of deriving an expression for
\beq
\mbox{lim}_{e \rightarrow \eta } \left[ e^a_{~~\alpha}
\funcderv{\Gamma ^{HTL(quark)} [\bar{\psi}, \psi, A, g]}{e^{a \beta }} \right]
\eeq
\noindent
where e is the vierbien field.

We start from a flat space angular integrand $X_{quark(0)}$
and action $\Gamma_{quark(0)}$ with a dependence on the  auxiliary fields
  $\chi
_{0}(Q, x)$ and $\Psi_{0}(Q, x)$ , the latter of which acts as a lagrange
multiplier (cf.  $V(Q, x)$ in section 2).
\beq \nonumber
&  g^2T^2 X_{quark(0)} =  \int d^{4}x a(\bar{\psi}\chi_{0} + \bar{\Psi_{0}}
(\slashchar{Q}\psi + iQ\cdot \tilde{D} \chi_{0})  + ~~~ h.conj.)
\\& \nonumber  \Gamma_{quark(0)} = ~g^2T^2 \int \frac{d\Omega}{4
\pi} X_{quark(0)}  &
\eeq
where $a=g^{2}T^{2}(N^2-1)/(16N)$ and $ \tilde{D}^{\lambda}$ is the
covariant derivative acting on fields
in the fundamental representation ie.  $ \tilde{D}^{\lambda} \psi =
\partial^\lambda + igA^\lambda \psi $.
The Euler-Lagrange equations give
\beq \label{PSIEQM}
 (Q\cdot \tilde{D})\chi_{0}=i\slashchar{Q} \psi  & \mbox{and} &
(Q\cdot \tilde{D})\Psi_{0}=i \psi
\eeq
Again, we impose sufficient restrictions on the fields belonging to
${\cal R}$ to ensure that (\ref{INTE}) holds for the non-local
functionals of the fields which we encounter in the quark sector. Thus we
have
\beq
 &\chi_0 = i (Q \cdot \tilde{D})^{-1} \slashchar{Q} \psi \hspace{1.2cm}
\Psi_0 = i
(Q \cdot \tilde{D})^{-1} \psi & \\
\label{CNQ}
 & \mbox{with} \spp \spp    (Q \cdot \tilde{D})^{-1}_{ret} \psi \spp =
\spp (Q \cdot
\tilde{D})^{-1}_{adv} \psi \spp \equiv \spp (Q \cdot \tilde{D})^{-1} \psi &
\eeq
where $(Q \cdot \tilde{D})^{-1}_{ret} $ and $ (Q \cdot
\tilde{D})^{-1}_{adv}$ are integral operators similar to
(\ref{INVERSE}) and given by
\beq
 ((Q \cdot \tilde{D})^{-1}_{ret} \psi)^A (x) &=&   \int_{- \infty}^0
d\theta  U(x, x+Q \theta ) \psi(x+Q \theta) \\ \nonumber
 ((Q \cdot \tilde{D})^{-1}_{adv} \psi)^A (x) &=& -  \int^{ \infty}_0
d\theta   U(x, x+Q \theta ) \psi(x+Q \theta)
 \\ \nonumber     \mbox{with} \qquad U(x, x+Q \theta ) &=& \Big(
     \mbox{P}      \exp  \int_{\theta}^0 d\theta ' [
                  -ig Q^\mu A_{\mu} (x+Q \theta ')] \Big)
\eeq

Thus we have on elimination of the auxiliary fields and using
 $[\slashchar{Q}, Q\cdot \tilde{D}]=0$,
\beq \label{qfsgf}
X_{(0)quark} = ia \bar{\psi}\slashchar{Q}(Q\cdot \tilde{D})^{-1}
\psi + \mbox{h.  conj. }
\eeq
\noindent
 This is the quark flatspace  hard thermal loop angular integrand denoted
$X^{HTL}_{quark(0)}$ and the above calculation is valid for fields in
${\cal R}$. We can consider (\ref{qfsgf}) for fields not in
${\cal R}$  by analytic continuation.

\subsection{Inclusion of weakly coupling graviton field to quark sector}

Again,  we consider generalising to curved space and
imposing Weyl invariance noting that  the Weyl transformation properties of
the auxiliary fields once more  do not affect the action at
$O(\kappa)$ (see Appendix B). Thus
$\Psi$ and $\chi$ can be considered, without loss of generality,  as Weyl
 invariants. Unlike
sections 2 and 3 we find no ambiguities on generalising the angular
integrand and
action. This is  consistent with the fact that quark sector analogues of the
gluon conditions (i) and (ii)
(given by (i)$'$ and (ii)$'$ in  section (A.3) of Appendix A) are
sufficient,without an
analogue of the gluon condition (iii) above, to determine
$X^{HTL(quark)}$. (See
section A.3 of Appendix A for further details). This  makes the quark
analysis relatively straightforward. $\Gamma^{HTL(quark)}$ and
$X^{HTL(quark)}$ are  defined analogously to their gluon counterparts ie.
\beq\nonumber
\ghtl^{(quark)} &=& \!\! \int \frac{d\Omega}{4 \pi}
d^4z_1 d^4z_2 \sum_{m, n} d^4x_{1}, \ldots, d^4x_{m}d^4y_{1},
\ldots, d^4y_{n}
 A_{\mu_1}(x_{1}), \ldots, A_{\mu_m}(x_{m}) \\ \nonumber
&& \hspace{-2.6cm} \varphi^{\alpha_1
\beta_1}(y_{1}), \ldots, \varphi^{\alpha_n \beta_n}(y_{n})
\bar{\psi}(z_1) G^{HTL(quark)}(x_{1}, \ldots, x_{n}, y_{1},
\ldots, y_{n};Q)^{\mu_1 \ldots
\mu_m}_{\alpha_1 \beta_1 \ldots \alpha_n \beta_n} \psi(z_2)\\  & \equiv &
g^2T^2 \int
\frac{d\Omega}{4 \pi} X^{HTL(quark)}(Q;\bar{\psi}, \psi, A, g]
\eeq
where $G^{HTL(quark)}$ refers to the hard thermal 1-loop truncated
diagram with a quark anti-quark pair of external fields, $n$
external  graviton fields and $m$ external gluon fields. As before we
take $n=1$ and assume that $X^{HTL(quark)}$ satisfies
conditions (A) $\rightarrow$ (D) given above.
Consider the Weyl scaling and general coordinate invariant
generalisation of the auxiliary field angular integrand in section (4.1) :
\beq \label{qlafa}
& g^2T^2 X_{quark} = \int d^{4}x \sqrt{-g}a
(\bar{\psi}e^{\frac{5\Lambda}{4}} \chi +
\bar{\Psi} ( e^{\frac{5\Lambda}{4}} (\yslash)\psi + i e^\Lambda
\dot{y} \cdot \tilde{\nabla} \chi) +
\mbox{h.  conj. }& \\ \nonumber
&\Gamma_{quark}  = \int \frac{d\Omega}{4 \pi} X_{quark} &
\eeq
\noindent
where
\beq
 (\dot{y} \cdot \tilde{\nabla}) \chi &=&
\dot{y}^\nu(\partial_\nu + igA_\nu + \omega'_{\nu bc} \sigma^{bc})
\chi \\
\mbox{with }~~~~~ \omega'_{\nu bc} &=& +\smallfrac{1}{2} e_b^{\rho}e_{c
\rho ; \nu} + \smallfrac{1}{4} (e_{b \nu} e_c^{ \rho} - e_{c \nu}e_b^\rho)
\Lambda ,  _\rho \\
\mbox{and }~~~  \yslash &=& e^{-\frac{\Lambda}{2}}(\dot{y}^{\nu}
e_{a \nu} \gamma^{a})
\eeq
Note that $e^\Lambda(\dot{y} \cdot \tilde{\nabla}) \chi$ and \yslash
 are  both  Weyl invariants.

We  now  eliminate the auxiliary fields  $\chi ,
\Psi$ from (\ref{qlafa}) and denote the resulting expression by
$X^*_{quark}$. We have from the Euler-Lagrange equation for $ \bar{\Psi }$
\beq \label{CSEM}
\funcderv{X_{quark}}{\bar{\Psi}} = 0 \quad \Rightarrow
(\dot{y} \cdot \tilde{\nabla} )\chi = i  [e^{\frac{\Lambda}{4}}(\yslash) \psi]
\eeq
Again, we restrict the fields so that  we
have equality between the retarded and advanced solutions of
(\ref{CSEM}). Thus,
\beq \nonumber
 ((\dot{y} \cdot \tilde{\nabla})^{-1} \chi)(x) &=& \int_{-\infty}^0
V(x, x+Q\theta ) \chi (x+Q \theta ) = \int^{\infty}_0
V(x, x+Q\theta ) \chi (x+Q \theta )
 \\  \label{zxc}
\mbox{where } ~~~~ V(x, x+Q \theta) &=&  \Big( \mbox{P}\exp -
            \int_{-\theta}^0  d\theta '  [\dot {y}^\nu
(igA_\nu(y(x,  \theta ')) +
\omega'_{\nu bc}(y(x,  \theta ')) \sigma^{bc})]  \Big)\\ \nonumber
            &=&  \Big( \mbox{P}\exp -
            \int_{-\theta'}^0  [ig \dot {y}^\nu
                A_\nu(y(x,  \theta '))]  \Big) \cdot \\
\nonumber && \hspace{1.5cm} \Big( \mbox{P}\exp -
            \int_{\theta '}^0 [\dot {y}^\nu \omega'_{\nu bc}(y(x,  \theta '))
\sigma^{bc}]  \Big)
\eeq
where $y(x,  \theta ')$ is given by (\ref{NGE}) subject to
(\ref{NGC}).

\noindent
 Noting that $\Psi$ and $\bar{\Psi}$ are
Lagrange multipliers, we can now eliminate  the auxiliary
fields. Using $[\yslash,   (\dot{y} \cdot \tilde{\nabla})^{-1} e^{\Lambda}]
 \chi =0 $ for any
$\chi$ (See Appendix D), we have the following expression for $X^*_{quark}$
\beq
X^*_{quark}=ia \int d^{4}x
\sqrt{-g}(\bar{\psi}e^{\frac{3\Lambda}{4}} \dot{\slashchar{y}}  (\dot{y} \cdot
\tilde{\nabla})^{-1} e^{\frac{\Lambda}{4}} \psi + \mbox{h.  conj. })
\eeq
It is straightforward to verify that $X^*_{quark}$ satisfies conditions
(i)$'$ and (ii)$'$ of section (A.3) of Appendix A, where it is shown
that this implies that $X^*_{quark} = X^{HTL(quark)}$ to $O(\kappa)$.
 Therefore, we can use the same methods as
before (section (2.3)) to calculate the quark energy
momentum tensor :
\beq
T^{HTL(quark)}_{\alpha \beta } &=& \int \frac{d\Omega}{4 \pi} S_{\alpha
\beta}^{quark} \\ \nonumber
\mbox{with } S_{\alpha \beta}^{quark} &\equiv& g^2T^2
\mbox{ lim}_{e \rightarrow
\eta} \Big( e^a_\alpha \funcderv{X^*_{quark}}{e^{a \beta}} \Big) =
g^2T^2 \mbox{ lim}_{e \rightarrow
\eta} \Big( e^a_\alpha \funcderv{X^{HTL(quark)}}{e^{a \beta}} \Big)
 \\ \nonumber
S_{\alpha \beta}^{quark} &=&  a \EMLAM [\pbzero \slashchar{Q}
\psi] \\ \nonumber
&+& a \EMYDOT [ \pbzero \gamma_{\lambda} \psi + i \pbzero D_{\lambda}
\slashchar{Q}
\pzero ] \\ \nonumber
&+& \smallfrac{i}{2}a[ \partial_{\mu}  \big( \bar{\Psi} Q_{\alpha}
\sigma _\beta ^{~~\mu} \chi + \bar{\Psi} Q_{\beta} \sigma _\alpha^{~~\mu}
\chi
+ \bar{\Psi} Q^\mu \sigma_{ \alpha \beta} \chi \big) ] + h. conj.
\eeq

\noindent

It is straightforward  verify that, up to $O(\kappa)$, $X^*_{quark}$
agrees with the quark
gluon graviton effective action given by equation (34) of
\cite{taylor}. They do not agree at $O(\kappa^2)$ despite the fact
that both satisfy conditions (i)$'$ and (ii)$'$ due to the fact that
our definitions of $\Lambda$ are different at $O(\kappa^2)$ (as
discussed further in Appendix C).

Again, these calculations are valid in ${\cal R}$ and may be
analytically continued to fields off ${\cal R}$.
\section{Conserved Currents}
First of all we note that for any $G_\lambda$
\beq \label{IDE}
   \int d^3x G_{0} = \int d^3x (Q \cdot \partial)^{-1} \partial^{\lambda}
G_{\lambda}
\eeq
 Using (\ref{IDE})
 it is straightforward to calculate the
integrated energy and momentum densities.
 Consider the gluon
sector. We have from (\ref{EMT3}) (with $E=0$)
\beq
P_{\alpha}  \equiv  \int d^3x T^{HTL}_{0 \alpha }  &=& \int
\frac{d\Omega}{4 \pi} d^3x 4 \mbox{tr}\Big[ (\smallfrac{1}{2}
\eta_{\alpha 0} \eta _{\mu \nu} -\eta_{\mu \alpha} \eta_{0 \nu}
)(  \smallfrac{3}{4}m^2_g W_{0}^\mu W_0^{\nu}) \\ \nonumber &&
\hspace{1.5cm} +  \smallfrac{1}{2}
Q_{\alpha} \eta_0^\lambda (V_{0 \mu}F^{\mu}_{~~\lambda} -
V_{0\mu}D_{\lambda}W^{\mu} + \partial_\mu(V_0^\mu W_{0 \lambda})) \Big]
\eeq
where $W_0^\mu$ and $V_0^\mu$ are given by (\ref{FSEQM}).
Notice that $P_{\alpha}$ is local before the elimination of the
auxiliary fields. It is straightforward to verify that this expression
agrees with $P_{\alpha}$ as would be calculated in other papers
(compare eg.  (3.13) of \cite{Weldon}) which Blaizot and Iancu \cite{bi},
using properties of the angular
integral $\int d\Omega (4 \pi)^{-1}$, have
shown to be equal to
\beq
P_{0}= \int \frac{d\Omega}{4 \pi} d^3x ~ \smallfrac{3}{4} m^2_g [(Q \cdot
D)^{-1}Q^\mu F_{0 \mu}][(Q \cdot D)^{-1}Q^\mu F_{0 \mu}]
\eeq
 Thus $P_0$ is in fact a positive definite functional
of the gluon fields even though this positivity is not manifest.

Similarly, we can write down $P^{quark}_{\alpha}$
\beq
 P^{quark}_{\alpha} &\equiv& \int d^3x T^{HTL(quark)}_{0 \alpha } \\
&=& \nonumber \int \frac{d\Omega}{4 \pi} \int d^3x a \Big[ -\eta_{\alpha 0}(
\bar{\Psi}_0 \slashchar{Q} \psi) + \bar{\Psi}_0 \gamma_{\alpha} \psi +
i \bar{\Psi}_0 D_\alpha \slashchar{Q} \Psi_0 \Big] + h. conj.
\eeq
where $\Psi_0$ is given by (\ref{PSIEQM}). We note that this expression is
a local function of the auxiliary
fields and does not appear to be positive definite which perhaps is to be
expected eg.  by comparison with the zero temperature quark
contribution to energy momentum tensor. $T^{HTL}_{\alpha \beta }$
and $T^{HTL(quark)}_{\alpha \beta }$  are symmetric and traceless (on use
of the equations of motion for $T^{HTL(quark)}$) (see \cite{wein}). Thus, the
construction of angular momentum and scaling currents/charges is
standard. They are defined by (for the gluon sector)
\beq \nonumber
M_{\mu \lambda \nu} = x_\mu (T^{HTL}_{\lambda \nu } - x_\nu
T^{HTL}_{\lambda \mu  }) && C_{\mu} = x^{\nu} T_{\mu \nu }^{HTL} \\
J_{\mu \nu} = \int d^3x M_{\mu 0 \nu} && C = \int d^3x C_{0}
\eeq
with similar definitions for the quark sector expressions.
It should be noted that these expressions unlike the $P_{\mu}$ are non
local even before the elimination of the auxiliary fields.

As a final exercise, we compare $P_0 = \int d^3x T^{HTL}_{00}$ with the
the Hamiltonian derived from the local auxiliary field  lagrangian density
${\cal L}$ where
 $g^2T^2 X_0(Q;m_0,A, \eta]  = \int d^4x {\cal L}$. We use the
$m_0^{\mu}$ field model for simplicity as it has less constraints then the
$V_0^\mu, W_0^\mu$ model where $V_0^\mu $ acts as a lagrange
multiplier.

In the following we write we write out colour indices $(A, B,
\ldots)$ instead of using a matrix notation. For clarity, we also drop the
 flat
 space label $_0$ although we are working with flat space fields in
this analysis. Thus all indices in the remainder of this section are either
colour or coordinate ones.
\beq
    {\cal L} = (2b)
[m^{\lambda A}Q^{\nu}F_{\lambda\nu}^A + \smallfrac{1}{2} (Q\cdot
Dm_{ \nu}^A)(Q\cdot Dm^{\nu A})] - \smallfrac{1}{4} F^A_{\mu \nu}F^{\mu
\nu A}
\eeq
The conjugate momemta for $m$ and $A$ are given respectively by
\beq
\Pi^{\mu A} = 2b (Q \cdot D)m^{\mu A} && \Pi_{gluon}^{\mu A} = E^{\mu A}
+ 2b(Q^\mu m^{0 A} - m^{\mu A})
\eeq
where $E^{\mu A} = F^{\mu A}_{~~0}$. Hence
\beq
 \partial_0 m^{\mu A} &=& \smallfrac{1}{2b} \Pi ^{\mu A} +
{\bf Q } \cdot {\bf D} m^{\mu A} + i f^{ABC} A_0^B
m^{\mu C}   \\ \nonumber
\partial_0 A^A_j &=& D_jA^A_0 - E_j^A \qquad \qquad \Pi_{gluon}^{0A}
\equiv 0 \\ \nonumber
{\cal H} &\equiv& \Pi_{\mu}^{A} \partial_0 m^{\mu A} +
\Pi_{gluon}^{\mu A} \partial_0 A^A_\mu - {\cal L} \\ \nonumber
 &=& \smallfrac{1}{2}({\bf E }^2 + {\bf B }^2) + \Pi^A_{\mu}
 (\Pi^{\mu A} + {\bf Q } \cdot {\bf D } m^{\mu A} +
i f^{ABC} A_0^B m^{\mu C}) \\ &-& 2b~m^{jA}Q^kF^A_{jk} + A_0^A
(D_j \Pi_{j}^A)
\eeq

 We want to compare $\int \frac{d\Omega}{4 \pi} \int d^3x {\cal H}$
with  $\int d^3x~ ^m S^{E'=0}_{00}$.

[Note that in deriving $^m S^{E'=0}_{\alpha
\beta}$ in (\ref{MTEN}) we have used the  equations of
motion for $m$ to drop a term with a $(Q^\mu m_\mu)$ dependence, which
we denote by $\delta (^mS^{E'=0}_{\alpha \beta})$
\begin{eqnarray*}
 \delta (^mS^{E'=0}_{\alpha \beta}) \!\! & =& \!\! -\smallfrac{1}{2}
\EMETA \\ &&
\hspace{7.0cm} [\partial_\mu (m^\mu Q^\rho (Q \cdot
Dm_\rho))  ]
\end{eqnarray*}
However it  is straightforward to check that $\int d^3x \delta
(^mS^{E'=0}_{00})=0$ and hence $\int d^3x ^m S^{E'=0}_{00}$ is given
by integrating (\ref{MTEN}) with $E'=0$].

Subtracting $\int \frac{d\Omega}{4 \pi} \int d^3x {\cal H}$  from $
\int d^3x ~^m S^{E'=0}_{00}$  we obtain
not zero but
\beq
\int \frac{d\Omega}{4 \pi} \int d^3x  2b~m^{0A}[ (Q \cdot D)^2 m^{0A} -
Q^{\mu}F^{0A}_{~~\mu} ] -  A_0^A(D_j \Pi_{j}^A + i
f^{BAC} \Pi_{\mu}^B m^{\mu C})
\eeq
Thus the $m$ field theory produces a non standard result due to the
unusual form of its `kinetic energy' term $\smallfrac{1}{2} (Q\cdot
Dm_{ \nu}^A)(Q\cdot Dm^{\nu A})$ which has  cross terms of time and
and spatial derivatives. However, the physical theory is obtained by
 elimination of
the auxiliary fields in which case the term proportional to $m^0$
drops out (see
\ref{FSMEQM}) leaving only  the Gauss constraint term, ie.  a gauge
transformation generator which is constrained to
zero  when considering the physical, gauge invariant, theory. Hence, we
see that for the physical theory  $\int \frac{d\Omega}{4 \pi} \int
d^3x {\cal H} = \int d^3x T^{HTL}_{00}$.


\section{Conclusions}
In the above we have shown how to derive the gluon and quark
contributions, $T^{HTL}_{\alpha \beta }$ and
$T^{HTL(quark)}_{\alpha \beta }$, to the retarded QGP energy momentum tensor
 using an auxiliary field method. These
tensors have some desirable properties :
\begin{enumerate}
      \item[a. ] They are automatically gauge invariant, without having
to add divergences `by hand'.
      \item[b. ] They are symmetric.  $T^{HTL}_{\alpha \beta } =
T^{HTL}_{\beta \alpha }$ from its definition  and
$T^{HTL(quark)}_{\alpha \beta } =  T^{HTL(quark)}_{\beta \alpha }$ as a
consequence of the local lorentz symmetry and the equations of motion
(See eg. \cite{wein}).
      \item[c. ] $T^{HTL}\quad \!\!\!\!  _{\alpha } ^{~~\alpha} =
 T^{HTL(quark)}
\quad \!\!\!\! _{\alpha} ^{~~\alpha} =0$ using Weyl scaling symmetry
 together with the equations of
motion.
      \item[d. ] $\partial^{\alpha}( T^{HTL}_{\alpha \beta } +
T^{HTL(quark)}_{\alpha \beta } )=0$, as required for a energy momentum
 tensor of a closed
system. This is a consequence of general
coordinate symmetry and the equations of motion.  (Note that we have
not introduced an external source in this paper and hence do not a
external source current on the right hand side of the above, unlike
Blaizot and Iancu).
      \item[e. ] Despite appearances, $T^{HTL}_{\alpha \beta }$ is a
positive definite
functional of the gluon fields, although to show this requires
properties of the angular integral ie.  $\int d\Omega (4 \pi)^{-1}$.
\end{enumerate}
Properties (a) $\rightarrow$ (c) allow straightforward construction
of the conserved quantities of the QGP ie.  momentum, angular momentum
and the scaling current as given above.

However, the expressions for $T^{HTL}_{\alpha \beta }$ and $
T^{HTL(quark)}_{\alpha \beta }$ are
cumbersome, being non-local functionals even before elimination of the
auxiliary fields. The angular momenta and scaling currents/charges are non
local even before elimination of the auxiliary fields making their
manipulation cumbersome too.
Finally we conclude by noting that the insights gained using the above
method have shown that $\Gamma^{HTL}[A, g]$ and
$\Gamma^{HTL(quark)}[\bar{\psi},\psi,A, g]$ are  more complicated than
previously thought and are  certainly not known at present to
$O(\kappa^2)$ .

\begin{appendix}

\section{Appendix A}
In this Appendix we discuss the result that any
expression, denoted $ X^{guess}(Q;A, g] $
which
\begin{enumerate}
       \item[i. ]  Satisfies conditions (A) $\rightarrow$ (D)
       \item[ii. ] Reduces to $X^{HTL}_{0}(Q;A, \eta]$ when $g\equiv \eta $
       \item[iii. ]Agrees with $X^{HTL} (Q;A, g]$ \quad at \quad $O(g^{2}
\kappa )$
\end{enumerate}
must in fact equal to $X^{HTL}(Q;A, g]$ at $O(\kappa)$ for all orders in
$g$. In appendix C we briefly discuss the
some of the further complications that arise at $O(\kappa^{2})$.
\newline
Firstly, we note that this result is much weaker than previous
speculation that (i) \& (ii) would be sufficient to imply equality
between $ X^{guess}(Q;A, g] $ and $ X^{HTL}(Q;A, g]$
\cite{taylor,taylor(a)}. As seen at the end of section (2.2.3) $X_{E}$
is a clear counterexample.

As condition (ii) gives equality at $O(\kappa^0)$ we are primarily
interested in the theory at $O(\kappa)$. It is
sufficient to consider
\beq
& S^{guess}_{\alpha \beta} \equiv 2 g^2 T^2
\mbox{lim}_{g \rightarrow \eta} \left(
\funcderv{ X^{guess}(Q;A, g] }{ g ^{ \alpha \beta}}   \right) &
\eeq
Suppose two different $ X^{guess}(Q;A, g] $
exist satisfying (i)$\rightarrow$(iii) above, leading to two different
$S^{guess}_{\alpha \beta}$ whose difference we denote by $ \Delta
S_{\alpha \beta}$.
\newline
We show below that (i)$\rightarrow$(iii) force $ \Delta
S_{\alpha \beta}$ to zero at all orders in $g$ which proves the
result stated at the start of this appendix.
\subsection{Conditions (i)$\rightarrow$(iii) $\Rightarrow
 \Delta S_{\alpha \beta}$=0}

Firstly note that (iii) implies that $ \Delta
S_{\alpha \beta}$ is zero at $O(g^{2})$. The idea of the proof at
higher orders in $g$  is
simple and in the same spirit as \cite{taylor(a)}. Write
down the most general $O(g^{s})$ [here $s \ge 3$] contribution to $ \Delta
S_{\alpha \beta}$ consistent with conditions (A)$\rightarrow$(C) of
(i) above. Next, in a step that requires a lot of tedious but not
difficult algebra,  show that the homogeneous Ward identities that correspond
to condition (D) of (i) above can only be satisfied if $ \Delta
S_{\alpha \beta}$ at this order is identically zero.

The most general $O(g^{s})$ contribution to $\Delta S_{\alpha
\beta}$, denoted $ \Delta
S_{\mu_{s+1} \mu_{s+2}}^{s} $, [where $\alpha \rightarrow \mu_{s+1}$
and $ \beta \rightarrow \mu_{s+2}$ ] can be written as the following
integral

\beq \label{DELT}
 \Delta S_{\mu_{ s+1} \mu_{s+2}}^{s}
&=& \int d^4y_{1} \ldots d^4y_{s}d^4p_{1}
\ldots d^4p_{s} e^{- \sum_{i=1}^{s} p_{i} \cdot( y_{i}-x) } \\
\nonumber && W^s _{\mu_{1} \ldots \mu_{s} \mu_{s+1} \mu_{
s+2}} A^{\mu_{ 1}}(y_1) \ldots A^{\mu_{ s}}(y_s)
\eeq
 where  $p_{i}$ $(i=1 \ldots s)$ denote the $i^{th}$ external gluon
momentum. $W^s$ is a
function of $p_{1} \ldots p_{s}$ given by the  expression
below. First, some definitions {\em (used only in this appendix)}.
\begin{eqnarray*}
  & Q_{\overbrace{\mu_{i} \mu_{j} }\ldots } \equiv Q_{\mu_{1}} \ldots
Q_{\mu_{i-1}}  Q_{\mu_{i+1}} \ldots Q_{\mu_{j-1}} Q_{\mu_{j+1}} \ldots
Q_{\mu_{s+2}} & \\
& A^{\ast}_{ijkl} \equiv A_{ijkl} \eta_{\mu_{i} \mu_{j} \mu_{k} \mu_{l}}
Q _{\overbrace{\mu_{i} \mu_{j} \mu_{k} \mu_{l}}} \qquad
B^{\ast}_{ij} \equiv
B_{ij} \eta_{\mu_{i} \mu_{j}} Q_{\overbrace{\mu_{i} \mu_{j}}} & \\ &
C^{\ast}_{ijk} \equiv (C_{ij})_{\mu_{k}} \eta_{\mu_{i} \mu_{j}}
Q_{\overbrace{\mu_{i} \mu_{j}}} &
\end{eqnarray*}
where $A_{ijkl}, B_{ij}, (C_{ij})_{\mu_{k}}$ are (thusfar arbitrary)
 coefficients constructed from the gluon momenta and their
contractions with $Q_{\mu}$ and each other. Note that the
$\mu_{k}$  index of $(C_{ij})_{\mu_{k}}$ corresponds to an index on some
combination of the
external momenta (and thus it does not belong to a $Q_{\mu}$ or $\eta$).
 ~$D_{\mu_{i}
\mu_{j}}$ and $E_{\mu_{i}}$ below are defined in the same way.
Thus $W^{s}$ can be written in form compatible with (A) $\rightarrow$
(C) as

\begin{eqnarray} \label{WEQN}
 W^{s}_{\mu_{1}} \ldots _{\mu_{s+2}} & = & \Big\{ (A^{\ast}_{1234} +
A^{\ast}_{1324} + A^{\ast}_{1423}) + (A^{\ast}_{1235} +
A^{\ast}_{1325} + A^{\ast}_{1523}) + \ldots
\\ \nonumber & + & (A^{\ast}_{s-1, s, s+1, s+2} +
A^{\ast}_{s-1, s+1, s, s+2} +   A^{\ast}_{s, s+2, s-1, s+13} )
 \\ \nonumber &+ & B^{\ast}_{12} +  \ldots + B^{\ast}_{s+1, s+2}
  +  (C^{\ast}_{123} + C^{\ast}_{231} + C^{\ast}_{321})
\\ \nonumber &+& \ldots
+ (C^{\ast}_{s-1, s, s+1} + C^{\ast}_{s, s+1, s-1} + C^{\ast}_{s+1, s, s-1})
+ D_{\mu_{1} \mu_{2}} Q_{\overbrace{\mu_{1} \mu_{2}}}
\\ \nonumber & +& \ldots
+D_{\mu_{s+1} \mu_{s+2}} Q_{\overbrace{\mu_{s+1} \mu_{s+2}}}   +
  E_{\mu_{1}}  Q_{\overbrace{\mu_{1}}}+ \ldots + E_{ \mu_{s+2}}
 Q_{\overbrace{\mu_{s+2}}} \Big\}
\end{eqnarray}

The above expression is over-general in that it does not take into
account all the symmetries of $W^{s}$, such as symmetry on interchange
 of graviton indices $(\mu_{s+1}
\leftrightarrow \mu_{s+2})$ and, from the definition of $W$, symmetry
under the interchange $(p_{\mu_i}, \mu_i) \leftrightarrow
(p_{\mu_j}, \mu_j)$.
These represent further constraints
on $A, \ldots, E$ (eg.  graviton indice symmetry requires
$(A_{ijk, s+1}=A_{ijk, s+2} \quad \&
\quad A_{i, s+1, j, s+2}=A_{i, s+2, j, s+1})$. However the Ward
identities are sufficient to force
$A, \ldots , E$ to zero even without such constraints and so these
constraints are not  explicitly incorporated into (\ref{WEQN})
for simplicity.
  We define
\beq
p_{s+1}^\mu = p_{s+2}^\mu = -\sum_{i=1}^{s} p_{i}^\mu
\eeq
and thus $p_{s+1}^\mu = p_{s+2}^\mu$ give the momentum of the external
graviton field.

We also define  $P_{i} =
p_{i} / (Q_{\mu} p^{\mu}_{i})$ for $i=1 \ldots
s+2$ and thus  $Q_{\mu}P^{\mu}_{i} =1$ for all $i$. Then the
 homogeneous Ward identities corresponding to gauge and
general coordinate invariance of condition (D) are given by
\beq \label{WDIX}
& P_{i}^{\mu_{i}} W_{\mu_{1} \ldots \mu_{i} \ldots \mu_{s+2}} = 0
\qquad i=1 \ldots s+2 &
\eeq
[Note that Weyl invariance implies a further homogeneous Ward identity
that must be satisfied (which is not required in this section
but will be used in the next subsection)]
\beq \label{WDID}
& \eta^{\mu_{s+1} \mu_{s+2}} W_{\mu_{1} \ldots \mu_{s+1} \mu_{s+2}} =
0 &
\eeq
Now apply the Ward identities given by (\ref{WDIX}) to
(\ref{WEQN}). First examine the
contribution which has (s-3) factors of  $Q_{\mu}$ in the contraction
of $P_{1}$
with $W$

\begin{eqnarray*}
  \big[ A_{2345} \eta_{\mu_2 \mu_3} \eta_{ \mu_4 \mu_5 } + A_{2435}
\eta_{\mu_2 \mu_4} \eta_{ \mu_3 \mu_5 } + A_{2534} \eta_{\mu_2 \mu_5}
\eta_{ \mu_3 \mu_4 } \big] Q_{\overbrace{\mu_{1} \mu_2 \mu_3 \mu_4 \mu_5
}}   &+&   \\
 \big[ A_{2346} \eta_{\mu_2 \mu_3} \eta_{ \mu_4 \mu_6 } + A_{2436}
\eta_{\mu_2 \mu_4} \eta_{ \mu_3 \mu_6 } + A_{2634} \eta_{\mu_2 \mu_6}
\eta_{ \mu_3 \mu_4 } \big] Q_{\overbrace{\mu_{1} \mu_2 \mu_3 \mu_4 \mu_6
}}  &+& \\ \ldots + \Big\{  \big[ A_{s-1, s, s+1, s+2} \eta_{\mu_{s-1}
\mu_{s}}  \eta_{ \mu_{s+1} \mu_{s+2} } + A_{s-1, s+1, s, s+2}
\eta_{\mu_{s-1} \mu_{s+1}} \eta_{ \mu_s \mu_{s+2} }  &+&
\\  A_{s-1, s+2, s, s+1} \eta_{\mu_{s-1} \mu_{s+2}}
\eta_{ \mu_s \mu_{s+1} } \big]
 Q_{\overbrace{\mu_{1} \mu_{s-1} \mu_{s} \mu_{s+1} \mu_{s+2}
}} \Big\}  & = & 0
\end{eqnarray*}

However all terms above correspond to independent tensors and hence
all the $A_{ijkl}$ in the above expression must be zero. Hence all
$A_{ijkl}$ with none of $\{ijkl\}$ equal to one must be
zero. Contracting with the other $P_{i}$ shows that all $A_{ijkl}$ must
be zero.
\newline
Now consider the contribution with (s-2) factors of  $Q_{\mu}$ in the
 contraction of
$P_{1}$ with $W$. This gives the $B_{ij}$ (with ${i, j}$ not equal to
one) in terms of linear combinations of the $A_{ijkl}$ which from
above are zero and  hence such $B_{ij}$ must be zero too. Explictly the
$P_{1}$ contraction gives

\begin{eqnarray*} \nonumber
  B_{23} \eta_{\mu_2 \mu_3} Q_{\overbrace{\mu_1 \mu_2 \mu_3}} +
  B_{24} \eta_{\mu_2 \mu_4} Q_{\overbrace{\mu_1 \mu_2 \mu_4}} + \ldots +
  B_{s+1, s+2} \eta_{\mu_s+1 \mu_s+2} Q_{\overbrace{\mu_{s+1} \mu_{s+2}
\mu_3}}  && \\
 + \mbox{ \bigg\{ Terms involving $A_{ijkl} =0$ from above \bigg\}
} = 0 &&
\end{eqnarray*}

Considering all $P_{i}$ contractions shows that all $B_{ij}$ are zero. By
examining contributions with increasing numbers of $Q_{\mu}$ it is
straightforward to show that all $(C_{ij})_{\mu_{k}}, D_{\mu_i \mu_j}$
and $E_{\mu_i}$ are zero as well, which proves the desired result.
Note that at $O(g^{2})$, ie.  the s=2term,  the above method is
invalidated because in order to  show the $A_{ijkl}$
are zero it is required to  consider terms with (s-3) factors of  $Q_{\mu}$.


We have now shown that the momentum space integrands of $X^{HTL}$ and
$X^{guess}$ are the same at $O(\kappa)$. However, performing the
momentum space integrals required to calculate $X^{HTL}$ and
$X^{guess}$ is only well  defined in Euclidean space for arbitrary fields.
 In Minkowski space the integrals are only defined for arbitrary
fields via analytic
continuation. However, for the fields
belonging to ${\cal R}$, the support of the fields' Fourier transforms
is sufficiently restricted  for the Minkowski momentum space integrals to
be well defined even without analytic continuation. Hence for fields
in ${\cal R}$ agreement of momentum space integrands gives equality
of  $X^{HTL}$ and $X^{guess}$. [For fields not in ${\cal R}$,  the
same analytic continuation would  have to be used for both  $X^{HTL}$ and
$X^{guess}$ to ensure  equality].

\subsection{Relaxing condition (iii) above}
We also briefly remark that if condition (iii) above is relaxed then
 $ \Delta S_{\alpha \beta}$ is not forced to zero by conditions (i) and
(ii). The most general form of  $W_{\mu_1 \mu_2 \mu_3 \mu_4}^{s=2}$
taking into account the further constraints mentioned above  is
 written down as $W_{\alpha \beta}^{\mu
\nu}(P, R, K)$ as given by equation (B1) of
\cite{taylor(a)} noting the following notational changes
\begin{eqnarray*}
  ( [P_1 , \mu_1] \sppp  ; \sppp [P_2, \mu_2] \sppp ; \sppp  [P_3 ,  \mu_3]
\sppp ; \sppp [P_4 ,  \mu_4] )
\leftrightarrow ( [R, \mu] \sppp ; \sppp [-P , \nu] \sppp  ;
 \sppp  [K,  \alpha]  \sppp ; \sppp [K,  \beta] )
\end{eqnarray*}
Of course, this agrees with (\ref{WEQN}) for s=2 on imposing the
additional constraints discussed above. Following the analysis of
 \cite{taylor(a)} we find that
equation (B2) of \cite{taylor(a)} holds but equation (B3) of
\cite{taylor(a)} is
erroneous. Writing all the equations required to explicity show this is
tedious and not very informative. However the end result , contrary to
the erroneous conclusion of \cite{taylor(a)}, is that the most general form
of the $O(g^2)$ contribution to $\Delta S(C)_{\alpha \beta}$ consistent
with the Ward identities is
given by

\beq \label{biquity}
\Delta S(C)_{\alpha\beta} = (gT)^{2} \int
d^4yd^4y'd^4pd^4r e^{-ip \cdot (y-x)+ir \cdot (y'-x)}A_{\mu}(y)
[C~ W_{\alpha\beta}^{\mu\nu}
(p, r)]A_{\nu}(y')
\eeq
with
\beq
  \lefteqn{ W_{\alpha\beta}^{\mu\nu}=\AMBIGUITY } \\
\nonumber & & \AMBIGUITYY \\ \nonumber &  &  \AMBIGUITYYY
\eeq

and C is either a constant or possibly a scalar dimensionless and
rational function of the $\{ (Q \cdot
 p), (Q \cdot r), (Q \cdot k) \}$ as multiplication by any other function
will contradict (A) $\rightarrow$ (D) of condition (i) above.

Note that the coefficient of (\ref{EMT3}) can be written in the form of
(\ref{biquity}) for $C=(Q\cdot k)^{2}/[(Q\cdot p)(Q\cdot
r)]$. This shows that the auxiliary field construction of
section 2, which for $E \neq 0$ satisfies (i) and (ii) but not
(iii), is consistent with the above. The same applies for section 3.

\subsection{The Quark Sector}
In this subsection we outline a proof of the following:
 any expression, denoted \\ $ X^{guess(quark)}(Q;\bar{\psi } , \psi ,A, e] $,
which
\begin{enumerate}
       \item[i. ]  Satisfies conditions (A) $\rightarrow$ (D)
       \item[ii. ] Reduces to $X^{HTL}_{0}[A, \eta]$ when $e\equiv \eta $
\end{enumerate}
must in fact equal to $X^{HTL(quark)}(Q; \bar{\psi} , \psi ,A, e]$ at
$O(\kappa)$ for all orders in
$g$ which {\em is } consistent with the assertions made in
\cite{taylor}. We proceed in the same way as above and first of all define
$S^{guess(quark)}_{\alpha \beta}$ by
\beq
& S^{guess(quark)}_{\alpha \beta} \equiv g^2 T^2 \mbox{lim}_{e \rightarrow
\eta} \left( e^a _\alpha
\funcderv{ X^{guess(quark)}(Q; \bar{\psi} , \psi , A, e] }{ e ^{a \beta}}
  \right) &
\eeq
Again, we suppose two different $ X^{guess}(Q;A, e] $
exist satisfying (i)$\rightarrow$(iii) above, leading to two different
$S^{guess}_{\alpha \beta}$ whose difference we denote by $ \Delta
S_{\alpha \beta}^{(quark)}$.
The most general form of the $O(g^s)$ contribution to
$ \Delta S_{\alpha \beta}^{(quark)}$ can be written as follows
(with $\alpha \rightarrow \mu_{s+1}$
and $ \beta \rightarrow \mu_{s+2}$)
\beq \label{DELTQ}
 \Delta S_{\mu_{ s+1} \mu_{s+2}}^{s} &=& \int d^4z_1 d^4z_2 d^4y_{1}
\ldots d^4y_{s+2} d^4r_1 d^4r_2 d^4p_{1} \ldots d^4p_{s} \\ \nonumber &&
\hspace{-2.8cm}
 e^{-\sum_{i=1}^{s} p_{i} \cdot ( y_{i}-x) - \sum_{j=1}^2 r_j \cdot (z_j-x) }
   \bar{\psi}(z_1) W^{s(quark)} _{\mu_{1} \ldots \mu_{s} \mu_{s+1}\mu_{s+2} }
               \psi(z_2)  A^{\mu_{ 1}}(y_1) \ldots A^{\mu_{ s}}(y_s)
\eeq
where $W^{s(quark)}$ is a function of $r_1, r_2, p_1, p_{s+2}$ which
are the momenta of, respectively, the anti-quark,quark and the $s$
gluon fields. $\mu_{s+1},\mu_{s+2}$ are the vierbien indices.
 The most general form of $W^{s=0(quark)}$ consistent with conditions
(A) $\rightarrow $ (C) is given by
\beq \nonumber
W^{s(quark)}_{\mu_{1} \ldots \mu_{s+2}} &=& Q_{\mu_1} \ldots
Q_{\mu_{s+2}}[A' + A \slashchar{Q}] + \sum _{j=1}^{s+2}
Q_{\overbrace{\mu_j}} [ B'_j \gamma_{\mu_j} + (B_{ \mu_j}^j + C_j
\gamma^{\mu_j}) \slashchar{Q}] \\ \label{genw}
&+& \sum_{s+2 \ge j > k \ge 1} Q_{\overbrace{\mu_j \mu_k}} [ (D_{jk}
\gamma_{\mu_j} \gamma_{\mu_k} + E_{jk} \eta_{\mu_j \mu_k}) \slashchar{Q}]
\eeq

where $A',A,B_j',C_j,D_{jk},E_{jk}$ are coefficients formed from the
various contractions of the external momenta with the following
possibilities: the external
momenta, the gamma
matrices or the $Q_\mu$.(Note that they contain no
$\slashchar{Q}$).The $B_{ \mu_j} ^j$  are proportional to
some linear combination of the external momenta. The exact from of the
above coefficients is heavily restricted by conditions $(A) \rightarrow (C)$.

Note that due to the relation $\gamma_{\mu_2} \gamma_{\mu_1} =
-\gamma_{\mu_1} \gamma_{\mu_2} + 2  \eta_{\mu_1 \mu_2} $ the most
general of form of $W^{s=0(quark)}$ can be written without terms
proportional to $\gamma_{\mu_j} \gamma_{\mu_k}$ with $j \le k$ and also
 with all
$\slashchar{Q}$ dependence written with a single $\slashchar{Q}$
exclusively on the right hand side of (\ref{genw}).

Now consider the
homogeneous Ward identities. First we define
\beq
p_{s+1}^\mu=p_{s+2}^\mu = - \Sigma _{i=1} ^s p_i^\mu -r_1^\mu -r_2^\mu
\eeq
and hence $p_{s+1}^\mu=p_{s+2}^\mu$ gives the momentum of the external
 vierbien field.
Then with the definition  $P_i^\mu = (Q \cdot p_i)^{-1} p_i^\mu$ for
$i=1 \ldots s+2$ the homogeneous Ward identities are given by
\beq \label{qwii}
P^{\mu_j}W^{s=0(quark)}_{\mu_1 \ldots \mu_j \ldots \mu_{s+2}}=0 && \mbox{
for } j=1 \ldots s+2
 \\
\eta^{\mu_{s+1} \mu_{s+2}}
W^{s=0(quark)}_{\mu_1 \ldots \mu_{s+1} \mu_{s+2}} =0 &&
W^{s(quark)}_{\mu_1 \ldots \mu_{s+1} \mu_{s+2}} = W^{s(quark)}_{\mu_1
\ldots \mu_{s+2} \mu_{s+1}}
\eeq
For the $s=0$ case, it is just an exercise in linear algebra to show
that these identities
force  $W^{s=0(quark)}_{\mu_1 \mu_2}$ to zero. Thus we have
the equality of $X^{HTL(quark)}$ and   $
X^{guess(quark)}$ at $O(g^2 \kappa)$ (or rather equality of their
momentum space integrands).

The method of determining equality at higher orders in $g$ is now
identical to the method of the gluon sector i.e. consider the Ward
identities given by (\ref{qwii}) and the coefficients of the terms
with increasing numbers of $Q^\mu$ factors. This is sufficient to
force  $W^{s(quark)}_{\mu_1 \ldots \mu_{s+2}}$ to zero as in the gluon
case.

Thus we have equality between the angular integrands of
$X^{HTL(quark)}$ and   $X^{guess(quark)}$ at $O(\kappa)$ for all
orders in $g$ and thus $X^{HTL(quark)} = X^{guess(quark)}$ at
$O(\kappa)$. [See remarks at the end of section (A.1)].
\section{Appendix B}

The generalisation of the effective action to include the interaction
of weakly coupling gravitons is independent (at $O(\kappa)$)
 of the Weyl scaling
properties of the auxiliary fields.
Suppose, as in section 2, we have the auxiliary fields $V(x, Q)$ and
$W(x, Q)$
transforming on Weyl scaling  as  $W_{\mu} \rightarrow e^{ 2u \sigma}
W_{\mu} $,  and $ V \rightarrow e^{ 2v \sigma } V_{\mu} $ with

Then, the Weyl invariant extension becomes
\beq
g^2T^2X_{E,u,v}  &\equiv& \int  \meas  2
\mbox{tr} \Bigg[ -e^{2u \Lambda} \WW
 \\ \nonumber &+& e^{-v \Lambda} V_{\mu}\dot{y}^{\lambda}F_{\nu\lambda}
 -V_{\nu}[\dot{y}^{\alpha}\nabla_{\alpha}(e^{-u \Lambda} W_{\mu}  +
e^{-u \Lambda} (W_{\nu}\nabla_{\mu}\dot{y}^{\nu} -
\dot{y}^{\alpha}W_{\alpha}\Lambda, _{\mu})] \\ \nonumber
 && \hspace{-1.5cm} +E \Big\{ e^{-(u+v)\Lambda } V_{\nu} [\ambig] \Big\}
\Bigg] \\ \nonumber \Gamma_{E} &\equiv & g^2T^2
\int \frac{d\Omega}{4 \pi} X_{E}(Q; V, W, A, g]
\eeq

Notice that at $O(\kappa )$ the coefficient of $E$ does not change as
$\Lambda \sim \nabla \dot{y} \sim O(\kappa )$.

Thus, we have
\beq \nonumber
 X_{E, u, v}(Q;V, W, A, g]  &=&
 X_{E, u=0, v=0}(Q;e^{-v \Lambda}V, e^{-u \Lambda}W, A, g]   \\
 \nonumber  &=& X_{E, 0, 0}(Q;V, W, A, g])   - \kappa
\int dx \Lambda ( uV \cdot
\funcderv{X}{V} + vW \cdot \funcderv {X }{W} ) + O(\kappa
^{2})
\eeq
Note that the $O(\kappa)$ term  on the right of the
above vanishes at solutions of the auxiliary field Euler-Lagrange
equations. Hence
on eliminating the auxiliary fields by use of the Euler-Lagrange
equations this term automatically drops out. Thus we  have equality
at $(O(\kappa))$ between the two angular integrals on elimination of
the auxiliary fields. This idea is easily applied to the other angular
integrands of sections 3 and 4 as well.

\section{Appendix C}

In this appendix we discuss further  ambiguities  that
occur  in attempting to write down $\Gamma^{HTL}[A, g]$ and
$\Gamma^{HTL (quark)}[A, e]$ to $O(\kappa^2)$ using their known
flatspace contributions and their symmetry properties given by (A)
$\rightarrow$ (D) above. Firstly consider
$\Lambda$, defined to be  a scalar which vanishes when $g = \eta$  with the
Weyl scaling property that $\Lambda\rightarrow \Lambda + 2 \sigma$ .

This does not fix $\Lambda$ uniquely at $O(\kappa^{2})$ and there are
at least two degrees of freedom in $\Lambda$ at this order.  Consider
$ \Lambda(a_{1}, a_{2}, x)$ defined by

\begin{eqnarray*}
  \Lambda(a_{1}, a_{2}, x) &=& - \int^{ \infty}_{0} d\theta
e^{\Lambda(a_{1}, a_{2}, y(x, \theta)) } \int^{ \infty}_{0} d\theta'
e^{ \Lambda(a_{1}, a_{2}, y(x, \theta')) }
\bigg\{ e^{-2 \Lambda(a_{1}, a_{2}, y(x, \theta'))}
\\ && \Big[ R_{\mu \nu}(y(x, \theta')) \dot{y}
^{\mu}(x,\theta') \dot{y}^{\nu}(x,\theta')
+ \smallfrac{1}{2}
\smallfrac{d \Lambda (a_{1}, a_{2}, y(x,  \theta '))}{d \theta '} ^{2}
+ a_{1} J_{1} + a_{2}
J_{2} \Big] \bigg\} \\
\mbox{where } \spp
 J_{1} &=& \smallfrac{1}{2} \nabla_{\alpha} \dot{y}^{\alpha}(x, \theta')
\nabla_{\beta} \dot{y}^{\beta}(x, \theta')
- \nabla_{\alpha} \dot{y}^{\beta}(x, \theta')
\nabla_{\beta} \dot{y}^{\alpha}(x, \theta') \\
J_{2}  &=& \nabla_{\beta} \dot{y}^{\alpha}(x, \theta')
\nabla^{\beta} \dot{y}_{\alpha}(x, \theta')
- \nabla_{\beta} \dot{y}^{\alpha}(x, \theta')
\nabla_{\alpha} \dot{y}^{\beta} (x, \theta')
\end{eqnarray*}

It is easily confirmed that that on Weyl scaling $J_{1}$ and $J_{2}$
both merely scale by $e^{4 \sigma}$ and  that $\int d\theta
e^{\Lambda (a_1,a_2,y(x,  \theta ))}$ is a Weyl invariant {\em if} $\Lambda
\rightarrow \Lambda + 2\sigma$ on Weyl scaling
 [as a null geodesic, although  invariant, is  reparameterised
on Weyl scaling with an affine
parameter, $\theta$,  transforming as $d\theta \rightarrow
e^{-2 \sigma (\theta)} d \theta ]$.

Thus, using the well known (see eg. \cite{hawk}) Weyl
 scaling properties of $R_{\mu \nu}$,  it
can be shown that on Weyl scaling
\beq
 \Lambda (a_1,a_2,x) \rightarrow \Lambda
(a_1,a_2,x) + 2\sigma (x)
\eeq
 for arbitrary $a_1$ and $a_2$. Thus we see that there is at least a
further 2-parameter ambiguity at $O(\kappa^2)$ in trying to
 determine  $\Gamma^{HTL}[A, g]$ and
$\Gamma^{HTL (quark)}[A, e]$.

Note that
$\Lambda(0, 0, x)$ is the same as the expression used for
 $\Lambda$ in \cite{taylor} [as given by  equation(21) on using
equation (55)] whereas $\Lambda(1, 0, x)$
 reduces to $\Lambda =
+\int^{ \infty}_{0} d \theta ' ~ \nabla_{\nu} \dot{y}^{\nu}
 \mid _{y(x, \theta')} $ ie.  the $\Lambda$ used
throughout this paper.

We also note that the local auxiliary field Lagrangians might well be
dependent at $O(\kappa^2)$ on the Weyl scaling properties of the
auxiliary fields suggesting the presence of further ambiguities
at $O(\kappa^2)$.

\section{Appendix D}
Here we show that for any $\chi$
\beq\label{AI}
[ e^{-\frac{\Lambda}{2}} \dot{\slashchar{y}}, (\dot{y} \cdot
\tilde{\nabla})^{-1} e^{\Lambda}] \chi =0
\eeq
where we recall that
\begin{eqnarray}
(\dot{y} \cdot \tilde{\nabla}) \chi &=&
\dot{y}^\nu(\partial_\nu + igA_\nu + \omega'_{\nu bc} \sigma^{bc}) \\
\mbox{with }~~~~~ \omega'_{\nu bc} &=& +\smallfrac{1}{2} e_b^{\rho}e_{c
\rho ; \nu} + \smallfrac{1}{4} (e_{b \nu} e_c^{ \rho} - e_{c \nu}e_b^\rho)
\Lambda ,  _\rho
\end{eqnarray}
It is sufficient to show that
\beq \label{AII}
[ e^{-\frac{\Lambda}{2}} \dot{\slashchar{y}}, (\dot{y} \cdot
\tilde{\nabla})] \chi =0
\eeq
(as $(\dot{y} \cdot \tilde{\nabla})$ is invertible,  we can write $\chi
=  (\dot{y} \cdot \tilde{\nabla})^{-1} \Psi$
and then operating on (\ref{AII}) by $(\dot{y}   \cdot
\tilde{\nabla})^{-1} e^{\Lambda}$ gives (\ref{AI}) for arbitrary $\Psi$.

Note that the following will be useful in going from the first to
the second line of the calculation below.
\beq \label{GEO}
[\gamma^a, \omega '_{\nu bc} \sigma^{bc}] = +2\omega ' _\nu
\hspace{0.3mm} ^a \hspace{0.3mm} _c
\gamma^c && e_{a \alpha} \dot{y}^\mu \partial_\mu \dot{y}^\alpha
= -\dot{y}^\mu
\dot{y}^\lambda \Gamma^{\alpha}_{\mu \lambda} e_{a \alpha}
\eeq
where the right hand equation is due to the geodesic equation for
 $\dot{y}^\mu$. Therefore

\begin{eqnarray*}
[ e^{-\frac{\Lambda}{2}} \dot{\slashchar{y}}, (\dot{y} \cdot
\tilde{\nabla})] \chi &=& [e^{-\frac{\Lambda}{2}}
\dot{y}^{\lambda}e_{a \lambda}\gamma^{a}, \dot{y}^\mu (\partial_\mu +
\omega'_{\nu bc} \sigma^{bc})] \chi  \\
 &=& e^{-\frac{\Lambda}{2}} \dot{y}^\mu \dot{y}^\lambda \gamma^c [
\Gamma^{\alpha}_{\mu \lambda} e_{c \lambda} + \smallfrac{1}{2}
\Lambda, _\mu e_{c \lambda} - e_{c \lambda}, \mu +2e_{a \lambda}
 \omega ' _\nu \hspace{0.3mm} ^a \hspace{0.3mm} _c \gamma^c ] \chi \\
 &=& e^{-\frac{\Lambda}{2}} \dot{y}^\mu \dot{y}^\lambda \gamma^c
[-(e_{c\lambda ; \mu} - 2e_{a \lambda} \{ +\smallfrac{1}{2}
e^{a \rho} e_{c \rho ; \mu} \} ) \\
&& \hspace{2.0cm} +\smallfrac{1}{2}(e_{c \lambda}
\Lambda , _\mu - e_{c \mu} \Lambda
, _\lambda + g_{\mu \lambda} e_c^\rho \Lambda ,  _\rho)]
\end{eqnarray*}
\noindent
Terms involving $\Lambda$ drop out (note that
$g_{\mu \lambda} \dot{y}^\mu \dot{y}^\lambda =0 $) and it is easy to
check that the remaining term is identically zero.

Hence $[ e^{-\frac{\Lambda}{2}} \dot{\slashchar{y}}, (\dot{y} \cdot
\tilde{\nabla})] \chi =0$ as required.

\end{appendix}

\vspace{1cm}
\noindent
{\bf Acknowledgements}\\
 I am grateful to John Taylor for many very helpful discussions. I
would also like to thank the PPARC for  support.

\end{document}